\documentclass[useAMS,usenatbib,usegraphicx]{mn2e}
\title[Families of dynamically hot stellar systems]{Families of dynamically hot stellar systems \\ over ten orders of magnitude in mass}
\author[I. Misgeld and M. Hilker]{I. Misgeld$^{1}$\thanks{E-mail: imisgeld@mpe.mpg.de} and 
        M. Hilker$^{1}$\thanks{E-mail: mhilker@eso.org}\\
$^{1}$European Southern Observatory, Karl-Schwarzschild-Strasse 2, 85748 Garching bei M\"unchen, Germany }

\begin{document}

\date{12 April 2011}

\pagerange{\pageref{firstpage}--\pageref{lastpage}} \pubyear{2011}

\maketitle

\label{firstpage}

\begin{abstract}
Dynamically hot stellar systems, whether star clusters or early-type galaxies, follow well-defined scaling relations over many orders of magnitudes in mass. These fundamental plane relations have been subject of several studies, which have been mostly confined to certain types of galaxies and/or star clusters so far. Here, we present a complete picture of hot stellar systems ranging from faint galaxies and star clusters of only a few hundred solar masses up to giant ellipticals (gEs) with $10^{12}$~M$_{\sun}$, in particular including -- for the first time -- large samples of compact ellipticals (cEs), ultra-compact dwarf galaxies (UCDs), dwarf ellipticals (dEs) of nearby galaxy clusters and Local Group ultra-faint dwarf spheroidals (dSphs). For all those stellar systems we show the effective radius--luminosity, effective radius--stellar mass, and effective mass surface density--stellar mass plane. Two clear families of hot stellar systems can be differentiated: the 'galaxian' family, ranging from gEs over Es and dEs to dSphs, and the 'star cluster' family, comprising globular clusters (GCs), UCDs and nuclear star clusters (NCs). Interestingly, massive ellipticals have a similar size--mass relation as cEs, UCDs and NCs, with a clear common boundary towards minimum sizes, which can be approximated by $R_{\mathrm{eff}} \geq 2.24 \cdot 10^{-6} \cdot M_{\star}^{4/5}$~pc. No object of either family is located in the 'zone of avoidance' beyond this limit. Even the majority of early-type galaxies at high redshift obeys this relation. The sizes of dEs and dSphs ($R_{\mathrm{eff}} \sim 1.0$~kpc) as well as GCs ($R_{\mathrm{eff}} \sim 3$~pc) barely vary with mass over several orders of magnitude. We use the constant galaxy sizes to derive the distances of several local galaxy clusters. The size gap between star clusters and dwarf galaxies gets filled in by low mass, resolving star clusters and the faintest dSphs at the low mass end, and by GCs/UCDs, NCs and cEs in the mass range $10^6<M_\star<10^9$~M$_{\sun}$. In the surface density--mass plane the sequences of star clusters and galaxies show the same slope, but are displaced with respect to each other by $10^3$ in mass and $10^2$ in surface density. Objects that fall in between both sequences include cEs, UCDs, NCs and ultra-faint dSphs. Both, galaxies and star clusters, do not exceed a surface density of $\Sigma_{\mathrm{eff}} = 3.17 \cdot 10^{10} \cdot M_{\star}^{-3/5}$~M$_{\sun}$~pc$^{-2}$, causing an orthogonal kink in the galaxy sequence for ellipticals more massive than $10^{11}$~M$_{\sun}$. The densest stellar systems (within their effective radius) are nuclear star clusters.
\end{abstract}


\section{Introduction}
Beginning with a pioneering paper by \citet{1973A&A....23..259B}, the well known Fundamental Plane (FP) relations have been used throughout the years by many authors to investigate global relationships among physical properties of stellar systems (mainly galaxies), such as surface brightness, absolute magnitude and physical size (e.g. \citealt{1977ApJ...218..333K, 1985ApJ...295...73K, 1987ApJ...313...59D}; \citealt*{1992ApJ...399..462B, 1993ApJ...411..153B}; \citealt{1997AJ....114.1365B, 2003AJ....125.1866B, 2007ApJ...654..897B, 2009ApJS..182..216K}, and many more).

However, some of these studies only focus on stellar systems of rather high luminosities/masses, excluding dwarf elliptical galaxies and globular clusters \citep[e.g.][]{1992ApJ...399..462B, 1993ApJ...411..153B, 2003AJ....125.1866B}. \citet{2007ApJ...654..897B} and \citet{2007ApJ...663..948G}, on the other hand, examine globular clusters and faint dwarf galaxies, which are again not considered in \citet{2009ApJS..182..216K}. In particular in early studies, there are considerable gaps in certain luminosity/mass ranges \citep[e.g.][]{1997AJ....114.1365B}. New, unusual stellar systems have been discovered in nearby galaxy clusters and the Local Group (LG) during the last decade, such as ultra-compact dwarf galaxies (UCDs), compact elliptical galaxies (cEs), and ultra-faint LG dwarf spheroidal galaxies (for references see Sect.~\ref{sec:sample} and Table~\ref{tab:specialobjects}). They have to be included in FP studies, in order to investigate possible relations to the conventional stellar systems.

It is thus of great interest to study FP relations with a sample of stellar systems covering the parameter space in luminosity, mass and physical size as complete as possible. With that in mind, we investigate the structural properties of various early-type, mostly gas-poor stellar systems. These dynamically hot stellar systems (i.e. stellar systems whose stars are on randomized orbits) span almost 25 orders of magnitude in luminosity, corresponding to 10 orders of magnitude in stellar mass, and 5 orders of magnitude in size. With up-to-date data on local galaxy cluster dwarf galaxies, ultra-faint LG dwarf spheroidals, cEs, UCDs and nuclear star clusters, this has not been shown before with such a complete coverage. Note that it is not our intention to present the correct sampling of the luminosity function of individual types of objects. This is the task of large imaging surveys (e.g. SDSS), or dedicated cluster surveys (e.g. ACSVCS).

\section{Sample description}
\label{sec:sample}

\begin{table}
	\caption{Absolute magnitudes $M_V$, effective radii $R_{\mathrm{eff}}$, and stellar masses $M_{\star}$ of compact elliptical galaxies, the compact object M59cO, and Local Group dwarf galaxies. The references are: (HL) \textsc{HyperLeda}, \texttt{http://leda.univ-lyon1.fr/};
		(1)~\citet{2008A&A...486...85C}; 
		(2)~\citet{2009ApJS..182..216K};
		(3)~\citet{1993ApJ...411..153B};
		(4)~\citet{2008MNRAS.391..685S};
		(5)~\citet{2005A&A...430L..25M};
		(6)~\citet{2008MNRAS.385L..83C};
		(7)~\citet*{2003AJ....125.1926G};
		(8)~\citet[][and references therein]{2007ApJ...663..948G};
		(9)~\citet{2007ApJ...656L..13I};
		(10)~\citet*{2008ApJ...684.1075M};
		(11)~\citet{2005MNRAS.363..918L};
		(12)~\citet{2009MNRAS.397.1748B};
		(13)~\citet{2010ApJ...710.1664D};
		(14)~\citet{2010ApJ...712L.103B};
		(15)~\citet{2006MNRAS.365.1263M};
		(16)~\citet[][and references therein]{2010ApJ...711..671K};
		(17)~\citet{2010MNRAS.407.2411C};
		(18)~\citet{2008ApJ...688.1009M};
		(19)~\citet{2009ApJ...705..758M};
		(20)~\citet{2010MNRAS.405L..11C};
		(21)~\citet{1998ARA&A..36..435M}. 
		$^{(*)}$ $M_V$ derived from $M_B$ with $B-V=0.96$~mag \citep{1995PASP..107..945F}. $^{(\dagger)}$ $M_{\star}$ derived from $B-V$ (see Sect.~\ref{sec:massestimate}). $^{(\dagger\dagger)}$ $M_{\star}$ derived from $g-i$ (see Sect.~\ref{sec:massestimate}).}
	\label{tab:specialobjects}
	
		\begin{tabular}{@{}l l r l l}
		\hline
		Name & $M_V$ & $R_{\mathrm{eff}}$ & $M_{\star}$ & Ref. \\
		~    & [mag] & [pc]               & [$M_{\sun}$] &      \\
		\hline
		A496cE       & $-18.0^*$      & 470 &  $5.8 \times 10^{9\;\dagger\dagger}$ & 1 \\
		NGC~4486B    & $-17.7$        & 198 &  $4.3 \times 10^{9\;\dagger}$    & 2, 3 \\
		NGC~5846A    & $-18.4$        & 517 &  $9.5 \times 10^{9\;\dagger}$    & 4, HL \\
		NGC~5846cE   & $-16.9^*$      & 291 & $2.2 \times 10^9$ & 20 \\
		CG$_{\mathrm{A1689,1}}$ & $-17.2$     & 370 & $4.3 \times 10^{9\;\dagger\dagger}$  & 5 \\
		CG$_{\mathrm{A1689,2}}$ & $-16.5$     & 225 & $2.3 \times 10^{9\;\dagger\dagger}$  & 5 \\
		M59cO        & $-13.0^*$      & 50  & $9.1 \times 10^7$ & 6 \\
		\hline
		Sagittarius  & $-15.0$    & 500 &  ...   & 7, 8 \\
		Sculptor     & $-9.8$     & 160 &  $1.8 \times 10^{6\;\dagger}$      & 7, 8, HL  \\
		Fornax       & $-13.1$    & 400 &  $2.8 \times 10^{7\;\dagger}$   & 8, 21  \\
		Leo I        & $-11.9$    & 330 &  $1.1 \times 10^{7\;\dagger}$   & 8, 21  \\
		Leo II       & $-9.8$     & 185 &  $1.4 \times 10^{6\;\dagger}$   & 8, 21  \\
		Sextans      & $-9.4$     & 630 &  ...   & 8  \\
		Carina       & $-9.3$     & 290 &  $1.1 \times 10^{6\;\dagger}$   & 8, HL  \\
		Ursa Minor   & $-8.9$     & 300 &  $1.9 \times 10^{6\;\dagger}$   & 8, 21  \\
		Leo T        & $-7.1$     & 170 &   $1.0 \times 10^5$  & 9  \\
		Canes Venatici I  & $-8.6$  & 564 & $3.0 \times 10^5$  & 10 \\
		Canes Venatici II & $-4.9$  & 74  & $8.0 \times 10^3$  & 10 \\
		Hercules          & $-6.6$  & 330 & $3.7 \times 10^4$  & 10 \\
		Coma Berenices    & $-4.1$  & 77  & $4.8 \times 10^3$  & 10 \\
		Bo\"otes I        & $-6.3$  & 242 & $3.4 \times 10^4$  & 10 \\
		Bo\"otes II       & $-2.7$  & 51  & $1.4 \times 10^3$  & 10 \\
		Ursa Major I      & $-5.5$  & 318 & $1.9 \times 10^4$  & 10 \\
		Ursa Major II     & $-4.2$  & 140 & $5.4 \times 10^3$  & 10 \\
		Willman I             & $-2.7$  & 25  & $1.5 \times 10^3$  & 10 \\
		Draco             & $-8.8$  & 221 & $3.2 \times 10^5$  & 10, 11 \\
		Segue I           & $-1.5$  & 29  & $6.0 \times 10^2$  & 10 \\
		Segue II          & $-2.5$  & 34  & ... & 12 \\
		Leo IV      & $-5.8$   & 206 &  ...    & 13 \\
		Leo V       & $-5.2$   & 133 &  ...    & 13 \\
		Pisces II   & $-5.0$   & 60  &  ...    & 14 \\
		Cetus       & $-11.3$  & 600 &  ...    & 15 \\
		Tucana      & $-9.5$   & 274 &  $1.1 \times 10^{6\;\dagger}$    & 16, 21 \\
		And I       & $-11.8$  & 682 &  $9.7 \times 10^{6\;\dagger}$    & 16, 21 \\
		And II      & $-12.6$  & 1248 & $2.3 \times 10^{7\;\dagger}$    & 16, HL \\
		And III     & $-10.2$  & 482 &  $2.5 \times 10^{6\;\dagger}$    & 16, HL \\
		And V       & $-9.6$   & 300 &  ...    & 15 \\
		And VI/Pegasus  & $-11.5$ & 420 & $6.3 \times 10^{6\;\dagger}$  & 15, 21 \\
		And VII     & $-13.3$  & 791 &  ...    & 16 \\
		And IX      & $-8.1$   & 552 &  ...    & 17 \\
		And X       & $-8.1$   & 339 &  ...    & 16 \\		
		And XI      & $-6.9$   & 145 &  ...    & 17 \\
		And XII     & $-6.4$   & 289 &  ...    & 17 \\
		And XIII    & $-6.7$   & 203 &  ...    & 17 \\
		And XIV     & $-8.3$   & 413 &  ...    & 16 \\
		\hline
		\end{tabular}
\end{table}

\begin{table}
\contcaption{}
\begin{tabular}{@{}l l r l l}
		\hline
		Name & $M_V$ & $R_{\mathrm{eff}}$ & $M_{\star}$ & Ref. \\
		~    & [mag] & [pc]               & [$M_{\sun}$] &      \\
		\hline
		And XVIII   & $-9.7$   & 363 &  ...    & 18 \\
		And XIX     & $-9.3$   & 1683 &  ...   & 18 \\		
		And XIX     & $-9.3$   & 1683 &  ...   & 18 \\		
		And XX      & $-6.3$   & 124 &  ...    & 18 \\
		And XXI     & $-9.9$   & 875 &  ...    & 19 \\
		And XXII    & $-6.5$   & 217 &  ...    & 19 \\
		\hline
		\end{tabular}
\end{table}

In this section, we describe in detail how we compiled the basic FP parameters effective radius $R_{\mathrm{eff}}$, absolute $V$-band magnitude $M_V$, and stellar mass $M_\star$. For the description of how $R_{\mathrm{eff}}$ was derived for the individual objects, we refer to the original source papers. The following objects are included in our study:

\begin{enumerate}
\item Giant elliptical galaxies, dwarf elliptical galaxies and bulges (including M32) from \citet{1993ApJ...411..153B}. For these objects, $M_V$ was calculated from the given absolute $B$-band magnitudes and the $B-V$ colours.

\item Early-type galaxies from the ACS Virgo Cluster Survey \citep[ACSVCS,][]{2006ApJS..164..334F}. The apparent $g_{\mathrm{AB}}$-band magnitudes were transformed into absolute $V$-band magnitudes using the relation $V = g_{\mathrm{AB}} + 0.026-0.307 \cdot (g-z)_{\mathrm{AB}}$ given in \citet{2006ApJ...639..838P}, and a Virgo distance modulus of $31.09$~mag \citep{2007ApJ...655..144M}. The same transformations were applied to a sample of nuclei of nucleated dwarf galaxies (dE,Ns) from the ACSVCS \citep{2006ApJS..165...57C}.

\item Bona fide extragalactic globular clusters (GCs) from the ACSVCS \citep{2009ApJS..180...54J}. $M_V$ was calculated in the same manner as for the ACSVCS galaxies. Note that in our study, $R_{\mathrm{eff}}$ of all objects from the ACSVCS is the average of the half-light radii measured in the $g$- and in the $z$-band.

\item Early-type galaxies (giant ellipticals and dwarf ellipticals) from the photometric studies of the galaxy clusters Hydra\,I and Centaurus \citep*{2008A&A...486..697M,2009A&A...496..683M}. We adopted a Hydra\,I distance modulus of $33.37$~mag, and a Centaurus distance modulus of $33.28$~mag, \citep*{2005A&A...438..103M}.

\item Compact elliptical galaxies identified in the HST/ACS Coma Cluster Treasury Survey \citep{2009MNRAS.397.1816P}. $M_V$ was estimated from the tabulated absolute $B$-band magnitudes, using $B-V=0.96$~mag \citep*{1995PASP..107..945F}.

\item Milky Way, LMC, SMC and Fornax star clusters from \citet{2005ApJS..161..304M}. The values for $M_V$ and $R_{\mathrm{eff}}$ were taken from the King models.

\item Ultra-compact dwarf galaxies with masses larger than $10^6$~M$_{\sun}$ from \citet{2008A&A...487..921M}, including the most massive UCDs in Fornax (UCD3) and Virgo (VUCD7). The tabulated masses from this study were converted into $M_V$ with the given $M/L_V$ ratios and a solar absolute magnitude of $M_{V,\sun} = 4.83$~mag \citep{1998gaas.book.....B}.

\item A sample of 22 nuclear star clusters (NCs) of spiral galaxies for which both the effective radii and the magnitudes in $V$ and $I$ are reported in \citet{2004AJ....127..105B} and \citet{2006AJ....132.1074R}.
\end{enumerate}

The photometric and structural parameters for the remaining objects, i.e. for the compact elliptical galaxies A496cE, NGC~4486B, NGC~5846A, NGC~5846cE, and for the two cEs in the galaxy cluster Abell~1689, for the compact object M59cO, and for the LG dwarf galaxies are listed in Table~\ref{tab:specialobjects}. $M_V$ and $R_{\mathrm{eff}}$ of NGC~5846A were calculated by adopting $V-R = 0.61$~mag \citep{1995PASP..107..945F}, and a distance modulus of $(m-M) = 32.08$~mag \citep[][and references therein]{2008MNRAS.391..685S}. The absolute magnitude of the LG dwarf galaxies, for which only the $V$-band luminosity $L_V$ is given in the source-paper, was calculated by $M_V=M_{V,\sun}-2.5\log(L_V/L_{\sun})$.

\subsection{Stellar mass estimates}
\label{sec:massestimate}
In order to estimate the stellar mass $M_\star$ of a particular object, we derived relations between its broad band colour and the stellar mass-to-light ratio $M/L_V$, using \citet{2005MNRAS.362..799M} simple stellar population (SSP) models.

For the objects from \citet{1993ApJ...411..153B}, we used a 13-Gyr SSP model, assuming a \citet{2001MNRAS.322..231K} initial mass function (IMF) and a red horizontal branch. The $M/L_V$--colour relation was parametrized as 
\begin{equation}
\label{eq:bvmodel_1}
\frac{M}{L_V} = 4.500 + 1.934 \cdot \arctan \left[ 8.464 \cdot \left((B-V) - 0.998\right) \right], 
\end{equation}
and is valid for $0.65<(B-V)<1.20$~mag. The stellar masses of the compact elliptical galaxies NGC~4486B and NGC~5846A were derived in the same manner (see Table~\ref{tab:specialobjects}).

The same SSP model, i.e. 13-Gyr, Kroupa IMF, red horizontal branch, was used for deriving the $M/L_V$ ratios of the early-type galaxies and GCs from the ACSVCS (Eq.~(\ref{eq:gzmodel_1})), and the Hydra\,I and Centaurus early-type galaxies (Eq.~(\ref{eq:vimodel})):
\begin{equation}
\label{eq:gzmodel_1}
\frac{M}{L_V} = 4.466 + 1.869 \cdot \arctan \left[ 4.385 \cdot \left((g-z)_{\mathrm{AB}} - 1.478\right) \right],
\end{equation}

\begin{equation}
\label{eq:vimodel}
\frac{M}{L_V} = 4.408 + 1.782 \cdot \arctan \left[ 11.367 \cdot \left((V-I) - 1.162\right) \right].
\end{equation}
These relations are valid for $0.80<(g-z)_{\mathrm{AB}}<1.90$~mag and $0.80<(V-I)<1.40$~mag, respectively.

For the galaxy nuclei from \citet{2006ApJS..165...57C}, we applied an 11-Gyr SSP model (Kroupa IMF, red horizontal branch), in order to account for the younger ages of those objects \citep[e.g.][and references therein]{2006ApJS..165...57C, 2009AN....330..969P, 2010MNRAS.405..800P}. The $M/L_V$ ratios are then given by
\begin{equation}
\label{eq:gzmodel_2}
\frac{M}{L_V} = 3.861 + 1.701 \cdot \arctan \left[ 3.795 \cdot \left((g-z)_{\mathrm{AB}} - 1.448\right) \right],
\end{equation}
for the colour range $0.80<(g-z)_{\mathrm{AB}}<1.90$~mag.

The stellar masses of 11 LG dwarf galaxies (see Table~\ref{tab:specialobjects}) were calculated from their $B-V$ colours, as given in \citet{1998ARA&A..36..435M} and \textsc{HyperLeda} \citep{2003A&A...412...45P}. For these objects, we derived a $M/L_V$--colour relation from an 11-Gyr SSP model (Kroupa IMF, blue horizontal branch), accounting for the, on average, younger ages and lower metallicities:
\begin{equation}
\label{eq:bvmodel_2}
\frac{M}{L_V} = 4.002 + 1.729 \cdot \arctan \left[ 7.619 \cdot \left((B-V) - 0.990\right) \right],
\end{equation}
valid for $0.65<(B-V)<1.20$~mag.

The stellar masses of the two cEs from \citet{2005A&A...430L..25M} were derived from the tabulated SDSS $(g-i)$ colours. First, these colours were transformed to $(V-I)$, using the relation $(V-I) = 0.671 \cdot (g-i) + 0.359$ from \citet*{2006A&A...460..339J}. Then, the stellar mass-to-light ratio was calculated with Eq.~(\ref{eq:vimodel}). The same method was applied to the compact elliptical A496cE from \citet{2008A&A...486...85C}, although here we used the $(g'-i')$ colours from the CFHT MegaCam filter system, which is however very similar to the SDSS filter system.

Accounting for the, on average, younger ages of nuclear star clusters \citep[][and references therein]{2006AJ....132.1074R}, we used an 8-Gyr SSP model (Kroupa IMF, red horizontal branch) to derive their $M/L_V$ ratios from the $V-I$ colours:
\begin{equation}
\label{eq:vimodel_2}
\frac{M}{L_V} = 2.913 + 1.269 \cdot \arctan \left[ 6.847 \cdot \left((V-I) - 1.120\right) \right].
\end{equation}
This relation is valid for $0.80<(V-I)<1.40$~mag.

\begin{table}
	\caption{Photometric and structural parameters of the objects used in this study. The references are: 
	    (1)~\citet{1993ApJ...411..153B};
		(2)~\citet{2006ApJS..164..334F};
		(3)~\citet{2008A&A...486..697M};
		(4)~\citet{2009A&A...496..683M};
		(5)~\citet{2008A&A...487..921M};
		(6)~\citet{2006ApJS..165...57C};
		(7)~\citet{2005ApJS..161..304M};
		(8)~\citet{2009ApJS..180...54J};
		(9)~\citet{2009MNRAS.397.1816P};
		(10)~\citet{2004AJ....127..105B} and \citet{2006AJ....132.1074R};
		(11)~see Table~\ref{tab:specialobjects} for detailed references.}
	\label{tab:fullcatalogue}
	
		\begin{tabular}{@{}l c c c c c}
		\hline
		ID & Ref. & $M_V$ & $\log(R_{\mathrm{eff}})$ & $\log(M_{\star})$ & $\log(\Sigma_{\mathrm{eff}})$ \\
		~  &  ~   & [mag] & [pc]                     & [$M_{\sun}$]      & [$M_{\sun}$~pc$^{-2}$]         \\
		\hline
        N0315 & 1 & $-24.6$ & 4.486 & 12.432 & 2.662 \\
        N0584 & 1 & $-22.6$ & 3.724 & 11.535 & 3.289 \\
        N0636 & 1 & $-21.6$ & 3.562 & 11.088 & 3.166 \\
        N0720 & 1 & $-22.6$ & 3.840 & 11.608 & 3.130 \\
        ... & & & & & \\
		\hline
		\end{tabular}
		
		\medskip
		This table is available in its entirety in a machine-readable form at the CDS. A portion is shown here for guidance regarding its form and content.
\end{table}

\begin{figure*}
	\includegraphics[width=168mm]{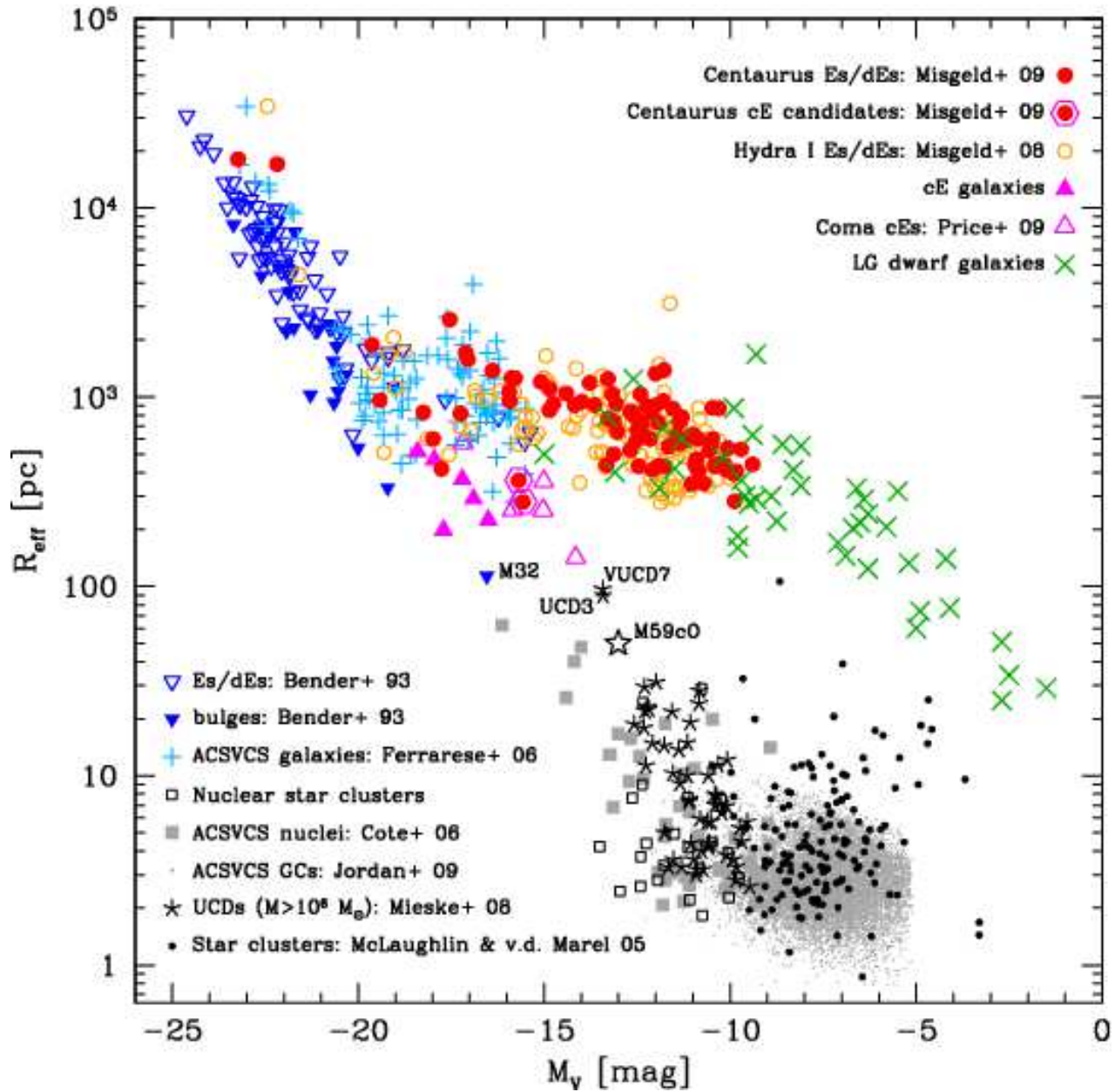}
	\caption{Effective radius $R_{\mathrm{eff}}$ plotted versus absolute $V$-band magnitude $M_V$ for the stellar systems described in Sect.~\ref{sec:sample}. Note the observational limit for the ACSVCS GCs at $M_V = -5$~mag.}
	\label{fig:sizelumidiag}
\end{figure*}

For comparison, we also computed stellar mass-to-light ratios for the above mentioned objects, using $M/L_V$--colour relations derived from \citet{2003MNRAS.344.1000B} SSP models (8-/11-/13-Gyr, Chabrier IMF). These models return smaller $M/L_V$ ratios, depending on the actual colour of the object. The percent differences\footnote{The percent difference of two values $x_1$ and $x_2$ is defined as $\mathrm{Diff} = \left|\frac{x_1-x_2}{(x_1+x_2)/2}\right|\times100$.} between the $M/L_V$ ratios of the two models are 1-10\% for Eq.~(\ref{eq:bvmodel_1}) and Eq.~(\ref{eq:bvmodel_2}), 11-25\% for Eq.~(\ref{eq:gzmodel_1}) and Eq.~(\ref{eq:gzmodel_2}), and 11-37\% for Eq.~(\ref{eq:vimodel}) and Eq.~(\ref{eq:vimodel_2}). The resulting differences in stellar mass, however, appear small in a logarithmic representation, and therefore do not change our conclusions. 

For the Coma cEs from \citet{2009MNRAS.397.1816P}, we computed $M_\star$ with the given $B$-band luminosities and stellar $M/L_B$ ratios, assuming a solar absolute $B$-band luminosity of $M_{B,\sun} = 5.48$~mag \citep{1998gaas.book.....B}.

The stellar masses of the Milky Way, LMC, SMC and Fornax star clusters (King model values) were directly taken from \citet{2005ApJS..161..304M}. 

We use the dynamical masses for the UCDs from \citet{2008A&A...487..921M}, since only for a few of them stellar masses can be derived. The differences between dynamical masses and stellar masses are smaller than 6\% of the UCD mass, and therefore do not change our results.

In Table~\ref{tab:fullcatalogue}, available at the CDS, we present the full catalogue of photometric and structural parameters of the objects used in this study. The first column is the object ID as given in the original source paper. A reference to the source paper is given in column 2. The third column lists the absolute $V$-band magnitude $M_V$. Columns 4, 5 and 6 give the logarithm of the effective radius $R_{\mathrm{eff}}$, the stellar mass $M_{\star}$, and the effective mass surface density $\Sigma_{\mathrm{eff}}$, respectively.

\section{Scaling relations}
\label{sec:fprelations}
Figure~\ref{fig:sizelumidiag} shows the $R_{\mathrm{eff}}$--$M_V$ plane for the objects discussed in Sect.~\ref{sec:sample}. In this plane, two distinct families or branches of objects can be identified. The first, 'galaxian', family comprises elliptical galaxies (giants and dwarfs), cEs and ultra-faint LG dwarf spheroidals, covering the full magnitude range of $-25 < M_V < -2$~mag (coloured symbols). The second family consists of 'star cluster-like' objects, i.e. GCs, UCDs, nuclei of dE,Ns and NCs (black and grey symbols). We emphasize, that this separation is only based on the morphological appearance of the objects, and does not imply that objects of one family have all formed by the same physical processes. 

At this point it is worth mentioning two different ways of dividing galaxies/star clusters into separate families. Based on luminosity, size and surface brightness, \citet{2009ApJS..182..216K} distinguished between a sequence of elliptical galaxies (ranging from typical giant elliptical galaxies to cEs like M32) and a sequence of spheroidal galaxies (see their fig.~38). Similarly, \citet{2008MNRAS.389.1924F} reported on a common sequence of giant elliptical galaxies, cEs and UCDs/GCs in a plot of virial mass ($\propto \sigma^2 R_h$) vs. stellar mass (their fig.~13), again with dwarf spheroidal galaxies deviating from this sequence \citep*[see also][]{2008MNRAS.386..864D}.

The second point of view is that (giant) elliptical galaxies and dwarf elliptical galaxies form a continuous sequence, extending from galaxies with a central light deficit to galaxies with a central light excess, based on the evaluation of the outer light profile by a S\'ersic law \citep[e.g.][]{2003AJ....125.2936G, 2006ApJS..164..334F, 2007ApJ...671.1456C}. In this picture, cEs and UCDs would be outliers of the galaxy sequence. In a similar manner, \citet{2011ApJ...726..108T} defined a one-dimensional fundamental curve through the mass-radius-luminosity space, connecting all spheroidal galaxies. Again, GCs and UCDs do not follow this fundamental curve relation. It is, however, not the scope of this paper to enter the discussion on which of those viewpoints is more appropriate.

\begin{table*}
	\caption{Results of the distance measurements. $a$ is the slope of the size--luminosity relation. References for the literature distance moduli $(m-M)_{\mathrm{lit}}$ are: $^{(a)}$~\citet{2005A&A...438..103M}; $^{(b)}$~\citet{2007ApJ...655..144M}; $^{(c)}$~\citet*{2003A&A...408..929D}; $^{(d)}$~\citet{2009ApJ...694..556B}.  }
	\label{tab:distind}
	\begin{tabular}{l c r r c c c c c}
	Sample & $a$ & $\langle r_{\mathrm{e}} \rangle$ & $\langle r_{\mathrm{e}} \rangle_{\mathrm{cor}}$ & $D$ & $D_{\mathrm{cor}}$ & $(m-M)$ & $(m-M)_{\mathrm{cor}}$ & $(m-M)_{\mathrm{lit}}$ \\
	~ & [arcsec mag$^{-1}$] & [arcsec] & [arcsec] & [Mpc] & [Mpc] & [mag] & [mag] & [mag] \\
	\hline
	Hydra I   & $-0.35 \pm 0.07$ & $ 3.58 \pm 0.19$ & $ 4.06 \pm 0.56$ & $57.6 \pm 6.5$ & $50.8 \pm 8.7$ & $33.80 \pm 0.57$ & $33.53 \pm 0.85$ & $33.37^a$ \\
	Centaurus & $-0.32 \pm 0.06$ & $ 4.25 \pm 0.25$ & $ 4.76 \pm 0.49$ & $48.5 \pm 5.7$ & $43.3 \pm 6.2$ & $33.43 \pm 0.58$ & $33.18 \pm 0.72$ & $33.28^a$ \\
	Virgo     & $-1.05 \pm 0.14$ & $11.08 \pm 0.42$ & $11.46 \pm 1.63$ & $18.6 \pm 2.0$ & $18.0 \pm 3.1$ & $31.35 \pm 0.53$ & $31.28 \pm 0.87$ & $31.09^b$ \\
	Antlia    & $-0.46 \pm 0.11$ & $ 5.49 \pm 0.24$ & $ 5.57 \pm 0.58$ & $37.6 \pm 4.1$ & $37.0 \pm 5.3$ & $32.88 \pm 0.55$ & $32.84 \pm 0.72$ & $32.73^c$ \\
	Fornax    & $-0.84 \pm 0.23$ & $ 7.47 \pm 0.45$ & $ 8.27 \pm 1.03$ & $27.6 \pm 3.2$ & $24.9 \pm 4.0$ & $32.20 \pm 0.58$ & $31.98 \pm 0.80$ & $31.51^d$ \\
	\hline
	\end{tabular}
\end{table*}

\subsection{Galaxies as distance indicators}
\label{sec:distind}

\begin{figure}
	\includegraphics[width=84mm]{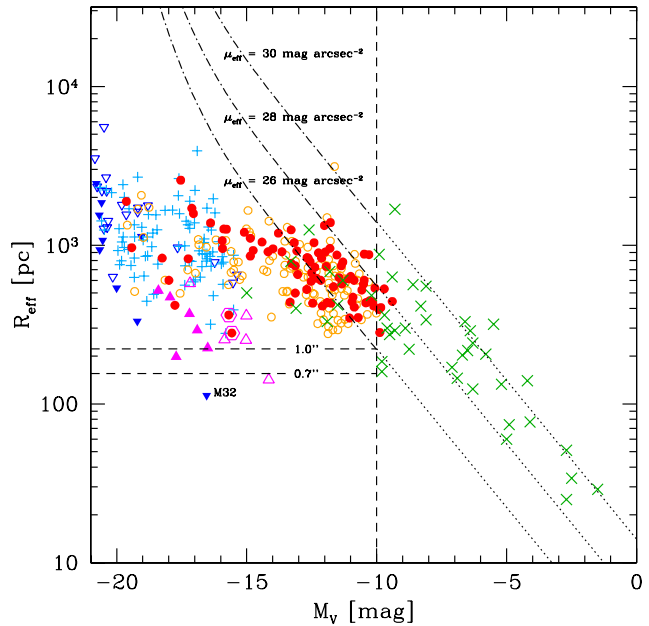}
	\caption{Enlargement of Fig.~\ref{fig:sizelumidiag} with various detection limits indicated. Only 'galaxy-like' objects are plotted, with the same symbols as in Fig.~\ref{fig:sizelumidiag}. The three (dash-)dotted lines indicate surface brightness limits of $\mu_{\mathrm{eff}} = 26, \, 28, \, 30$~mag~arcsec$^{-2}$ (see text for more details). The horizontal dashed lines represent resolution limits of $0.7''$ and $1.0''$, respectively, at the distance of Hydra\,I and Centaurus \citep{2005A&A...438..103M}. The vertical dashed line marks the 50\% completeness limit for the Hydra\,I and Centaurus dwarf galaxies at $M_V \sim -10$~mag \citep{2008A&A...486..697M, 2009A&A...496..683M}.}
	\label{fig:selectioneffects}
\end{figure}

An interesting feature in Fig.~\ref{fig:sizelumidiag} is is the nearly constant mean effective radius of $R_{\mathrm{eff}} \sim 1$~kpc for galaxies with $-21 \la M_V \la -10$~mag, which is equivalent to about 5 orders of magnitude in stellar mass (cf. Fig.~\ref{fig:sizemassdiag}). Equation~(3) in \citet{2009A&A...496..683M} quantifies the relation between $\log(R_{\mathrm{eff}})$ and $M_V$ for this particular magnitude range. In order to determine to what extend the constancy of $R_{\mathrm{eff}}$ and its downturn at fainter magnitudes is affected by selection effects, we show in Fig.~\ref{fig:selectioneffects} an enlargement of Fig.~\ref{fig:sizelumidiag}, with various detection limits indicated. 

The (dash-)dotted lines mark surface brightness limits of $\mu_{\mathrm{eff}} = 26, \, 28, \, 30$~mag~arcsec$^{-2}$. For calculating these limits, we used equations~(9) and (12) from \citet{2005PASA...22..118G}, approximating $\langle \mu \rangle_{\mathrm{e,abs}} \approx \langle \mu \rangle_{\mathrm{e}}$. As $\langle \mu \rangle_{\mathrm{e}}$ is a function of the S\'ersic index $n$ (equations~(8) and (9) in \citealt{2005PASA...22..118G}), we set $n=1$ for $M_V>-10$~mag. For brighter magnitudes, we fitted a 3rd order polynomial to the Hydra\,I and Centaurus data \citep{2008A&A...486..697M, 2009A&A...496..683M}, to describe $n$ as a function of $M_V$. The SDSS detection limit is $\sim 30$~mag~arcsec$^{-2}$ \citep{2008ApJ...686..279K}. The surface brightness limit in the Hydra\,I/Centaurus data is $27$--$28$~mag~arcsec$^{-2}$. Also indicated in Fig.~\ref{fig:selectioneffects} are typical seeing limits of $0.7''$ and $1.0''$ at the distance of Hydra\,I and Centaurus, as well as the 50\% completeness limit for the Hydra\,I and Centaurus dwarf galaxies at $M_V \sim -10$~mag.

At low luminosities ($M_V>-10$~mag) the trend of $R_{\mathrm{eff}}$ with $M_V$ is clearly affected by the SDSS detection limit for LG dwarf spheroidals, causing a steeper slope of the $\log(R_{\mathrm{eff}})$--$M_V$ relation \citep[see also][]{2009A&A...496..683M}. Very extended, low surface brightness objects cannot be detected at these magnitudes. For the Hydra\,I/Centaurus dwarf galaxies this is only true close to the 50\% completeness limit, as indicated in Fig.~\ref{fig:selectioneffects}. However, the inclusion of objects with large effective radii and low surface brightnesses, which could have potentially been missed at these luminosities, would rather lead to a flatter $\log(R_{\mathrm{eff}})$--$M_V$ relation. Further objects that could have been missed due to limited image resolution, are very compact, M32-like galaxies. These objects, however, fall below the main body of normal elliptical galaxies and do not bias the $\log(R_{\mathrm{eff}})$--$M_V$ relation.

\begin{figure}
	\includegraphics[width=84mm]{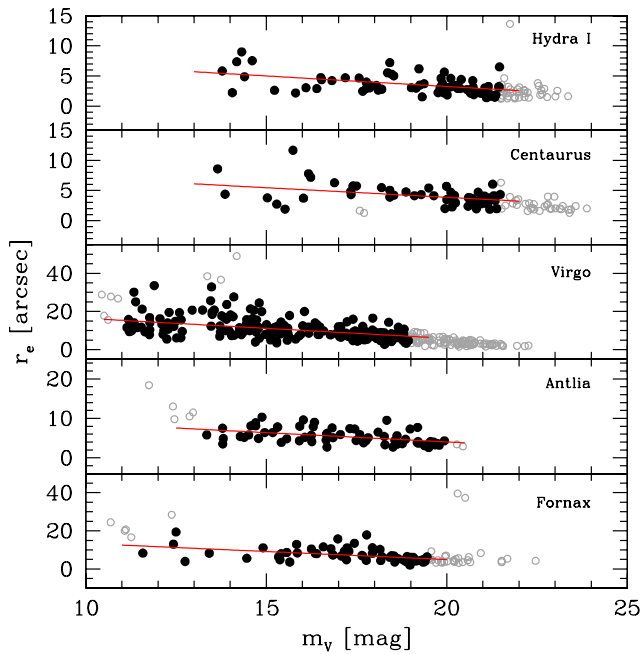}
	\caption{Size--luminosity relations in nearby galaxy clusters. The effective radius $r_{\mathrm{e}}$ is plotted against the apparent magnitude $m_V$. Filled black circles are the galaxies used for the measurements of the slope of the size--luminosity relation (as indicated by the red solid lines), and the mean effective radius (see also Table~\ref{tab:distind}). Grey open circles denote the rejected objects.}
	\label{fig:distind}
\end{figure}

We conclude that the nearly constant mean effective radius, which is observed over a wide range of magnitudes is not caused by selection biases, but is rather a genuine phenomenon. If confirmed in more galaxy clusters, this feature could serve as a distance indicator, provided that one can accurately determine the structural parameters of galaxies over a wide range of magnitudes. First efforts in this direction have already been made \citep[e.g.][and references therein]{2008MNRAS.386.2311S}.

We further investigate the potential of this distance indicator in Fig.~\ref{fig:distind}. Plotted is the apparent effective radius vs. the apparent magnitude of early-type galaxies in Hydra\,I \citep{2008A&A...486..697M}, Centaurus \citep{2009A&A...496..683M}, Virgo (\citealt{2006ApJS..164..334F}; Lieder et~al. 2011, in prep.), Antlia \citep{2008MNRAS.386.2311S}, and Fornax (\citealt*{2003A&A...397L...9H}; \citealt{2007A&A...463..503M}). For each sample, we measured $a$, the slope of the size--luminosity relation, and $\langle r_{\mathrm{e}} \rangle$, the mean effective radius. In order not to be affected by the selection effects discussed before, we restricted each sample to a magnitude range of $-20 \la M_V \la -12$~mag, according to the distance moduli given in the literature. Additionally, obvious outliers and the two cEs in Centaurus were excluded (grey symbols in Fig.~\ref{fig:distind}). The $V$-band magnitude of the Antlia galaxies was calculated, using eq.~(2) from \citet{2008MNRAS.386.2311S}. The slope $a$ was determined by fitting a linear relation to the data, applying a $3\sigma$ rejection algorithm. The mean effective radius $\langle r_{\mathrm{e}} \rangle$ was used to calculate the cluster distance $D$, employing
\begin{equation}
D = \frac{d}{\delta},
\end{equation} 
in which $\delta$ is the apparent mean effective radius $\langle r_{\mathrm{e}} \rangle$ in angular units, and $d$ is the true mean effective radius in pc, for which we assumed a value of $d=1.0 \pm 0.1$~kpc (the exact mean value measured in Fig.~\ref{fig:sizelumidiag} in the magnitude range $-20 < M_V < -12$~mag is $d=982$~pc).

Due to the negative slope of the size--luminosity relation, and due to the shape of the galaxy luminosity function, the mean effective radius $\langle r_{\mathrm{e}} \rangle$ is biased towards smaller values, caused by more data points with lower values of $r_{\mathrm{e}}$ at faint magnitudes. In order to correct for this bias, we subdivided each sample into bins of $2$~mag width, determined the mean effective radius of each bin, and defined the average of those values as the final, corrected $\langle r_{\mathrm{e}} \rangle_{\mathrm{cor}}$.

Table~\ref{tab:distind} summarizes the results of the measurements. Column~2 gives the slope $a$ of the size--luminosity relation, columns~3--8 list both the uncorrected and corrected values of the mean effective radius in arcsec, the resulting distance in Mpc, and the according distance modulus. For comparison, the literature distance moduli are given in the last column.

In all clusters considered, the slope of the size--luminosity relation is shallow. For Hydra\,I, Centaurus and Antlia the measured values agree within the errors, only in Virgo and Fornax the slope is slightly steeper. This justifies the assumption of an almost constant effective radius over a wide range of galaxy luminosities and different cluster environments. The derived distance moduli are well in agreement with the reported literature values (within about $0.2$~mag), although the values have rather large errors, caused by the scatter in the observed effective radii and the slope of the size--luminosity relation. Only for Fornax we measure a distance modulus deviating $\sim +0.5$~mag from the literature value. This might be caused by having only a few data points available at magnitudes brighter than $m_V = 15$~mag (see Fig.~\ref{fig:distind}), leading to an overall smaller mean effective radius an thus to a larger distance modulus. Note that at intermediate and low luminosities, the scatter of the size--luminosity relation might artificially be reduced by the non-detection of both very extended low-surface brightness objects and very compact objects. The latter (e.g. cEs) are, however, rare compared to the number of regular dwarf elliptical galaxies, and have thus been excluded from our analyses. At the lowest luminosities considered ($M_V \sim -12$~mag), very extended objects might have been missed, but the comparison with LG dwarf galaxies shows that the number of such non-detections should be small (cf. Fig.~\ref{fig:selectioneffects}).

In summary, the use of galaxy mean effective radii seems to offer a reasonable alternative to estimate the cluster distance, given that it is possible to identify the suitable magnitude range to perform the measurements. On the one hand, one has to avoid the magnitude regime where the steep size--luminosity relation of giant elliptical galaxies sets in, and on the other hand, one has to take care of not being affected by surface brightness limitations at faint luminosities.

\subsection{The sizes of hot stellar systems}
\label{sec:sizes}
Although having a large range of luminosities in common ($-15 \la M_V \la -5$~mag), dwarf galaxies and star clusters/UCDs are well separated in size, the latter being approximately two orders of magnitude smaller. This has previously been noted by \citet{2007ApJ...663..948G}, stating that there are no stable objects in a size gap between $\sim 30$~pc and  $\sim 120$~pc. However, Fig.~\ref{fig:sizelumidiag} shows that with several UCDs, compact elliptical galaxies, very extended star clusters and ultra-faint LG dwarf galaxies, this size gap is not as well-defined as it appeared in \citet{2007ApJ...663..948G}, in particular at bright ($M_V \sim -15$~mag) and very faint ($M_V \ga -5$~mag) magnitudes. Some of the faint star clusters and dwarf spheroidal galaxies are certainly in an unstable stage of disruption or evaporation, and will therefore not reside at their position in the diagram for a very long time. Whether this is also true for very bright and massive objects in this size range, like M32, UCD3, VUCD7 or M59cO, remains an open question. These objects might originate from larger and more luminous galaxies, now being transformed by tidal interactions with a major host galaxy \citep[e.g.][]{2001ApJ...557L..39B, 2003MNRAS.344..399B, 2003Natur.423..519D}. Yet, there remains a prominent, 'hole-like' region in between the galaxy- and the star cluster-branch, with only very few objects therein.

Another interesting feature in Fig.~\ref{fig:sizelumidiag} is the steep size--luminosity relation for giant elliptical galaxies and bulges above a magnitude of $M_V \sim -20$~mag. Surprisingly, a similar relation is visible for cEs, the nuclei of dE,Ns, UCDs and NCs. Thus, there is a boundary, in the sense that the effective radius of a stellar system of a given luminosity is larger than $\log(R_{\mathrm{eff}}) = -0.33 \cdot M_V - 3.90$. The relation for the compact systems, however, sets in at much lower magnitudes of $M_V \sim -10$~mag. Below this limit the star cluster sizes are largely independent of magnitude, just like the sizes of (dwarf) elliptical galaxies below $M_V \sim -20$~mag.

\begin{figure*}
	\includegraphics[width=168mm]{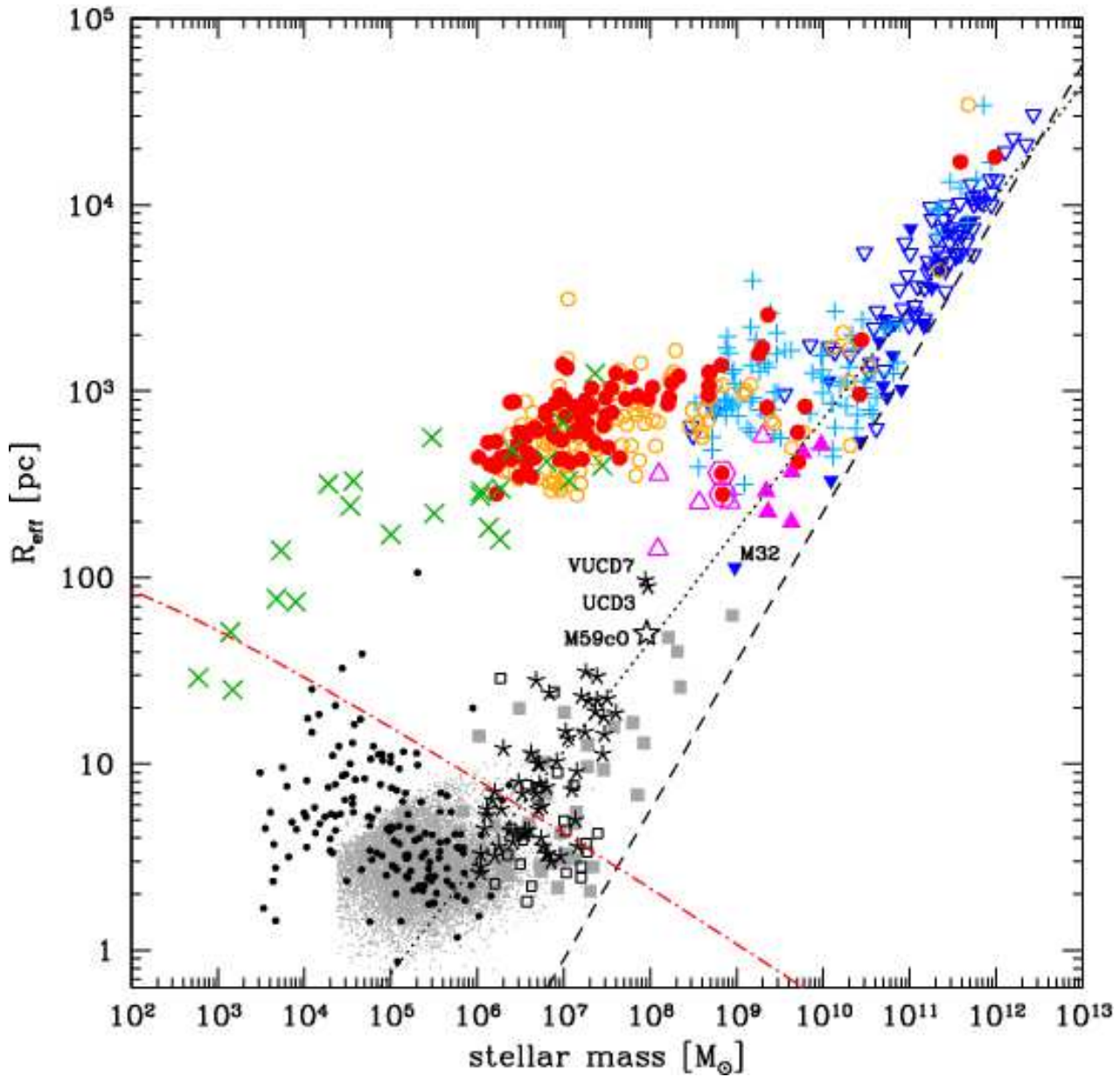}
	\caption{Effective radius $R_{\mathrm{eff}}$ plotted versus stellar mass $M_{\star}$ for all objects, for which $M_{\star}$ was computed as outlined in Sect.~\ref{sec:massestimate}. Symbols are as in Fig.~\ref{fig:sizelumidiag}. The dashed line represents Eq.~(\ref{eq:sizemass}), the dotted line is the size--mass relation from \citet{2008MNRAS.386..864D}. The (red) dash-dotted line shows Eq.~(\ref{eq:dab08}) for $t_{\mathrm{rel}}$ equal to a Hubble time.}
	\label{fig:sizemassdiag}
\end{figure*}

The boundary remains visible when looking at the size--mass plane in Fig.~\ref{fig:sizemassdiag}. Even more, giant elliptical galaxies and nuclei of dE,Ns define the same sharp edge in the diagram, i.e. there is a maximum stellar mass for a given effective radius. No object, whether star cluster or galaxy, is located beyond this limit, i.e. there is a 'zone of avoidance' in the parameter space. In $\kappa$-space, this has already been noted in \citet{1997AJ....114.1365B} as the 'zone of exclusion'. The boundary is illustrated by the dashed line in Fig.~\ref{fig:sizemassdiag}, which was adapted by eye to the data. In other words, the effective radius of an object of a given mass is always larger than:
\begin{equation}
\label{eq:sizemass}
R_{\mathrm{eff}}(M) \geq c_1 \cdot M_{\star}^{4/5},
\end{equation}
with $c_1=2.24 \cdot 10^{-6}$~pc~M$_\odot^{-4/5}$. In between the giant elliptical  galaxies and the nuclei of dE,Ns, i.e. in the mass interval $10^9 \la M_{\star} \la 10^{10}$~M$_\odot$, several compact elliptical galaxies, including M32, can be found close to this edge. \citet{2008MNRAS.386..864D} already mentioned a common size--mass relation for UCDs and giant elliptical galaxies \citep[see also fig.~6 in][]{2009MNRAS.397..488P}. The slope these authors found for their fitted relation is proportional to $M^{3/5}$ (see dotted line in Fig.~\ref{fig:sizemassdiag}). 

\citet{2009ApJ...691..946M} predicted a mass--radius relation $r_{cl} \propto M_{cl}^{3/5}$ for star clusters with an initial mass $M_{cl} \ga 3 \times 10^6$~M$_\odot$, which were supported by radiation pressure during their formation process. The difference between the size--mass relation $\propto M^{3/5}$ and Eq.~(\ref{eq:sizemass}) can be explained by including NCs and the nuclei of dE,Ns, which are at the same mass up to ten times smaller than UCDs \citep[see also][]{2008AJ....136..461E}. These objects might have already formed more compact than isolated star clusters via recurrent gas inflow, which causes repeated star formation events \citep[e.g.][]{2005ApJ...618..237W, 2006AJ....132.1074R}.

It is interesting to mention the study of \citet{2010MNRAS.408L..16G}, stating that low-mass GCs might have formed with the same size--mass relation as their more massive counterparts, and have until the present day moved away from this relation because of dynamical evolution. \citet{2008MNRAS.386..864D} give a formula for the median two-body relaxation time $t_{\mathrm{rel}}$ in stellar systems, which depends on the mass $M$ and the half-light radius $r_{\mathrm{e}}$ of the system (their eq.~(6)). Rearranging this equation gives

\begin{equation}
\label{eq:dab08}
r_{\mathrm{e}} = \left[ \frac{G \left(\log\left(M/M_{\odot}\right) \cdot t_{\mathrm{rel}} \right)^2}{0.0548\,M} \right]^{1/3},
\end{equation}

with $G=0.0045$~pc$^3$~M$_\odot^{-1}$~Myr$^{-2}$. In Fig.~\ref{fig:sizemassdiag} we plot this relation for $t_{\mathrm{rel}}$ equal to a Hubble time (red dash-dotted line). Objects below this line can thus have undergone considerable dynamical evolution since their formation, provided that they are old objects. It turns out that almost all low-mass GCs ($M_{\star} \la 10^6$~M$_\odot$) fall below this line, and that in the mass interval $10^6 \la M_{\star} \la 10^7$~M$_\odot$ the line divides very well objects which show a size-mass relation from objects which do not show such a relation (see also Fig.~\ref{fig:surfacedensity}). This supports the picture of low-mass GCs being considerably affected by dynamical evolution, as outlined in \citet{2010MNRAS.408L..16G}. Interestingly, also the ultra-faint dwarf spheroidals Segue~I and Willman~I fall below this line. They are suspected to be objects out of dynamical equilibrium and close to disruption, rather than ordinary dwarf spheroidal galaxies \citep{2007ApJ...663..948G, 2009MNRAS.398.1771N}.

\subsection{Galaxies and their star cluster mates}
\begin{figure*}
	\includegraphics[width=168mm]{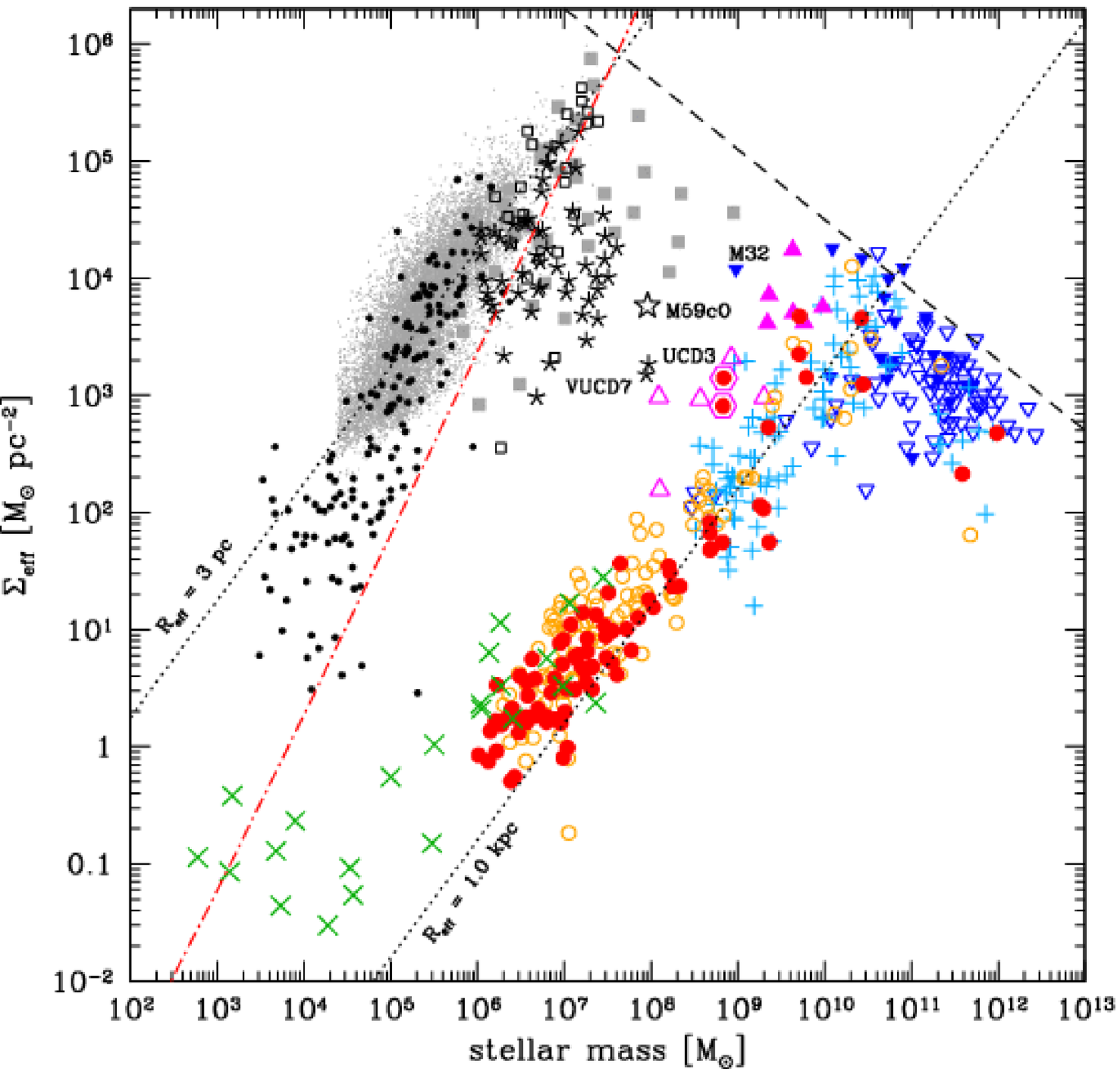}
	\caption{Mass surface density averaged over the projected effective radius, $\Sigma_{\mathrm{eff}} = M_{\star}/2\pi R_{\mathrm{eff}}^2$, plotted versus stellar mass $M_{\star}$ for all objects from Fig.~\ref{fig:sizemassdiag}. The dashed line represents Eq.~(\ref{eq:thresdensity}), the dotted lines mark the loci of objects with a constant radius of $R_{\mathrm{eff}} = 1.0$~kpc and $R_{\mathrm{eff}} = 3$~pc. The (red) dash-dotted line shows Eq.~(\ref{eq:sigmadef}) with $R_{\mathrm{eff}}$ as given by Eq.~(\ref{eq:dab08}) and $t_{\mathrm{rel}}$ equal to a Hubble time.}
	\label{fig:surfacedensity}
\end{figure*}

Although galaxies and star clusters occupy different locations in the $R_{\mathrm{eff}}$--$M_V$ plane, they still show intriguing parallels, like the size--mass relation which sets in above a certain mass limit (see Sect.~\ref{sec:sizes}). The similarities become even more obvious when plotting the mass surface density averaged over the projected effective radius, 
\begin{equation}
\label{eq:sigmadef}
\Sigma_{\mathrm{eff}} = M_{\star}/2\pi R_{\mathrm{eff}}^2, 
\end{equation}
versus the stellar mass $M_\star$, as shown in Fig.~\ref{fig:surfacedensity}. For both star clusters and galaxies, a sequence of increasing mass surface density with increasing stellar mass becomes visible. The stellar systems with the highest mass surface densities (up to $7.5 \times 10^5$~M$_{\sun}$~pc$^{-2}$) are NCs and nuclei of dE,Ns. However, they are embedded in spiral galaxies and dwarf galaxies, respectively. The densest \textit{isolated} systems are the most massive GCs and UCDs. 

Both sequences are close to identical, except for an offset of $\sim 10^3$ in mass and $\sim 10^2$ in surface density. Above a certain stellar mass, a kink towards lower mass surface densities appears. Galaxies with $M_{\star} \ga 5 \times 10^{10}$~M$_\odot$ as well as star clusters above $M_{\star} \sim 2 \times 10^{6}$~M$_\odot$ are arranged almost orthogonally to the main sequences. This results in a mass dependent maximum mass surface density, which is a consequence of the size--mass relation found in Fig.~\ref{fig:sizemassdiag}. Combining Eq.~(\ref{eq:sizemass}) and Eq.~(\ref{eq:sigmadef}) yields
\begin{equation}
\label{eq:thresdensity}
\Sigma_{\mathrm{eff}}(M) \leq c_2 \cdot M_{\star}^{-3/5},
\end{equation}
with $c_2=3.17 \cdot 10^{10}$~pc$^{-2}$~M$_{\sun}^{8/5}$. For comparison, \citet{2005ApJ...618..237W} find the empty top right region of Fig.~\ref{fig:surfacedensity} to be confined by $\Sigma_{\mathrm{eff}} \propto M^{-1/2}$, very similar to what we find. Note, however, that they plot $\Sigma_{\mathrm{eff}}$ against the dynamical mass rather than the stellar mass.

Again, in the $\Sigma_{\mathrm{eff}}$--$M_\star$ plane most cE galaxies reside offset to the main body of elliptical galaxies towards higher mass surface densities at a given stellar mass. This is caused by a mean size difference of $\sim 730$~pc between cEs (including M32) and regular elliptical galaxies in the mass range of the cEs ($10^8 \la M_{\star} \la 2 \times 10^{10}$~M$_\odot$). However, the largest cEs share the locus of normal elliptical galaxies (see also Fig.~\ref{fig:sizelumidiag} and Fig.~\ref{fig:sizemassdiag}). Since cEs like M32 are suspected to be the result of galaxy stripping processes \citep[e.g.][]{1973ApJ...179..423F, 2001ApJ...557L..39B}, one would have to investigate the true formation history of each questionable object in order to decide whether it is a compact elliptical galaxy having experienced intense tidal stripping, or simply a rather small genuine elliptical galaxy. Based on their observed broad-band colours, many cEs are found redwards the colour-magnitude relation of regular cluster early-type galaxies \citep[e.g.][]{2009A&A...496..683M, 2009MNRAS.397.1816P}. This can be interpreted as support for the stripping scenario, if the progenitor galaxy was a more luminous/massive galaxy, obeying the colour-magnitude relation. 

The red dash-dotted line in Fig.~\ref{fig:surfacedensity} shows Eq.~(\ref{eq:sigmadef}) with $R_{\mathrm{eff}}$ as given by Eq.~(\ref{eq:dab08}) and a median two-body relaxation time $t_{\mathrm{rel}}$ equal to a Hubble time. Thus, it divides the $\Sigma_{\mathrm{eff}}$--$M_\star$ plane into an area to the left of the line, in which objects have a two-body relaxation time shorter than a Hubble time (and thus their shape parameters might be changed within a Hubble time), and into an area in which $t_{\mathrm{rel}}$ is longer than a Hubble time (to the right of the line). \citet{1998MNRAS.300..200K} postulated that star cluster-like objects with masses $\la 10^9$~M$_\odot$ and $t_{\mathrm{rel}}$ longer than a Hubble time, evolved from massive stellar superclusters, which were created during gas-rich galaxy mergers \citep[see also][]{2002MNRAS.330..642F}. Figure~\ref{fig:surfacedensity} now suggests that UCDs with masses $\ga 2\times 10^6$~M$_\odot$ could indeed be the descendants of such stellar superclusters. The most massive UCDs are in fact observed in high density regions, i.e. galaxy clusters, where the central giant elliptical galaxies most probably formed via violent dissipative processes, like intense starbursts or early, gas-rich galaxy mergers. Interestingly, massive young clusters (with $M_{\star} > 10^{7}$~M$_\odot$) in starburst and merging galaxies already show a size--mass relation similar to those of UCDs \citep*{2006A&A...448.1031K}.

\subsection{Zone of avoidance}
\label{sec:zone}

\begin{figure}
	\includegraphics[width=84mm]{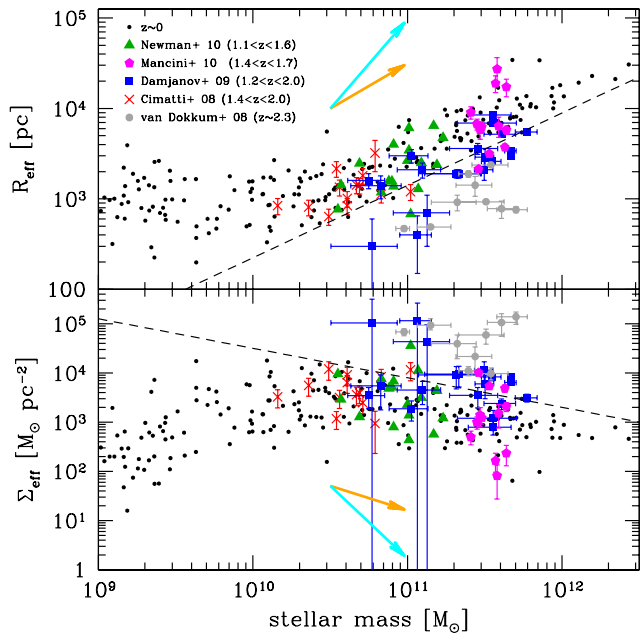}
	\caption{Enlargement of Fig.~\ref{fig:sizemassdiag} (upper panel) and Fig.~\ref{fig:surfacedensity} (lower panel). Only 'galaxy-like' objects with $M_{\star}> 10^9$~M$_{\sun}$ are plotted (black dots). The coloured symbols denote high redshift early-type galaxies. The orange (cyan) vectors indicate the size/surface density evolution according to the virial theorem for major (minor) galaxy mergers. The length of the vectors indicates a mass increase by a factor of 3.}
	\label{fig:limit}
\end{figure}

The sharply defined maximum stellar mass for a given half-light radius (Eq.~(\ref{eq:sizemass})), which translates into a maximum mass surface density (Eq.~(\ref{eq:thresdensity})), is evident for giant elliptical galaxies as well as for cEs, UCDs, nuclei of dE,Ns and NCs. No galaxy in the local Universe is found in the 'zone of avoidance' beyond this boundary \citep[see also][]{1997AJ....114.1365B}. This raises the question of whether it is coincidence or a physical law that causes this phenomenon.

In Fig.~\ref{fig:limit}, we compare the distribution of local early-type galaxies (ETGs) and high-redshift ($1.1 \la z \la 2.3$) ETGs in the $R_{\mathrm{eff}}$--$M_\star$ and the $\Sigma_{\mathrm{eff}}$--$M_\star$ planes. The high-$z$ ETGs are taken from \citet{2008A&A...482...21C}, \citet{2008ApJ...677L...5V}, \citet{2009ApJ...695..101D}, \citet{2010MNRAS.401..933M} and \citet{2010ApJ...717L.103N}, and have stellar masses of $2\times 10^{10} \la M_\star \la 5 \times 10^{11}$~M$_\odot$. The position of most of the high-$z$ ETGs is fully consistent with their $z \sim 0$ counterparts, i.e. also the high-redshift galaxies are not located beyond the critical boundaries given by Eqs.~(\ref{eq:sizemass}) and (\ref{eq:thresdensity}).

An exception is the sample from \citet{2008ApJ...677L...5V}. At a given stellar mass these ETGs have significantly smaller effective radii, corresponding to effective surface densities being several times higher than those of local ETGs. However, \citet{2010MNRAS.401..933M} showed that the sizes of high-$z$ ETGs can be underestimated by up to a factor of 3 at low S/N, preventing the detection of extended low surface brightness profiles, which are typical for massive elliptical galaxies. The underestimation of the profile shape index $n$ can also lead to a wrong determination of the effective radius $R_{\mathrm{eff}}$ \citep[see][]{2009MNRAS.398..898H}. Moreover, the stellar mass estimates decisively depend on the choice of the stellar population model. The ETG masses in the \citet{2008ApJ...677L...5V} sample are derived from \citet{2003MNRAS.344.1000B} models, thus, they are $\approx$~40--50\% higher than masses derived from \citet{2005MNRAS.362..799M} models \citep{2006ApJ...652...85M, 2008A&A...482...21C, 2010MNRAS.401..933M}. Given these possible sources of size underestimation and mass overestimation, respectively, the \citet{2008ApJ...677L...5V} ETGs might still be consistent with the $z\sim 0$ objects.

\begin{figure}
	\includegraphics[width=84mm]{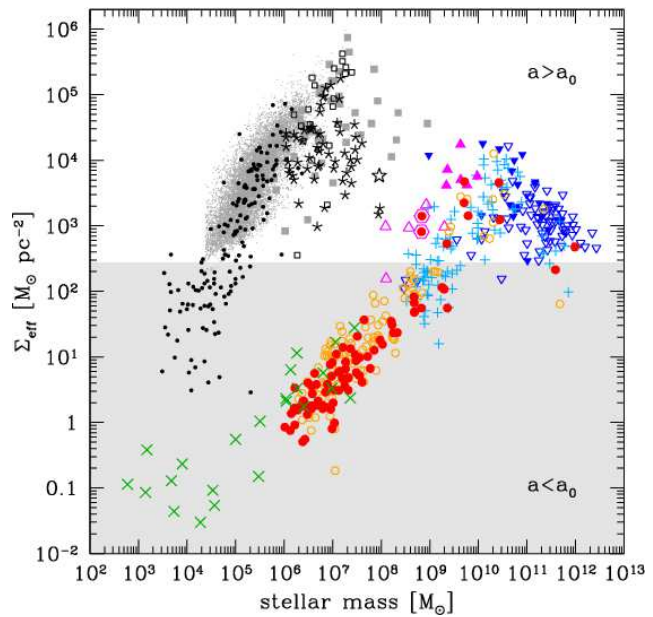}
	\caption{Same as Fig.~\ref{fig:surfacedensity}, but with the region shaded, in which a star in the stellar system experiences at the effective radius an acceleration smaller than $a_0 = 3.9$~pc~Myr$^{-2}$.}
	\label{fig:mond}
\end{figure}

Based on the virial theorem it can be shown that the radius $r$ of an evolving ETG increases linearly with stellar mass in the case of major mergers, and as the square of the mass in the case of minor mergers \citep*[e.g.][]{2009ApJ...699L.178N}. Indeed, there is observational evidence that the most massive ellipticals follow a linear one-to-one relation between $r$ and $M_{\star}$ \citep{2010MNRAS.tmp.1737T}. Regardless of which process, major or minor mergers, is mainly responsible for their subsequent mass assembly, the high-$z$ ETGs will move approximately along the boundary ($r \propto M_\star$), or away from it ($r \propto M_{\star}^2$), but they will not move across the limit (see Fig.~\ref{fig:limit}). 

The slope of the size--mass relation of the local ETGs is consistent with major merger evolution. The scatter might be caused by the effects of minor mergers. Even very compact high-$z$ ETGs \citep[e.g.][]{2008ApJ...677L...5V} can end up on the size--mass relation of $z \sim 0$ galaxies via successive major and minor mergers (cf. Fig.~\ref{fig:limit}). This reinforces the interpretation of the 'zone of avoidance' for $z \sim 0$ ETGs as the result of galaxy evolution.

\subsection{Internal accelerations}
The acceleration a star experiences at the effective radius inside a pressure-supported system is
\begin{equation}
\label{eq:accel}
a_{\mathrm{eff}} = \frac{GM}{R_{\mathrm{eff}}^2},
\end{equation}
where $M=0.5\cdot M_\star$ is the stellar mass within $R_{\mathrm{eff}}$. Combining this with Eq.~(\ref{eq:sigmadef}) yields the mass surface density as a function of $a_{\mathrm{eff}}$:
\begin{equation}
\label{eq:sigma_a}
\Sigma(a_{\mathrm{eff}})= \frac{a_{\mathrm{eff}}}{\pi G}.
\end{equation}
Setting $a_{\mathrm{eff}} = a_0 = 3.9$~pc~Myr$^{-2}$, which is the critical acceleration parameter in the theory of modified Newtonian dynamics \citep[MOND,][]{1983ApJ...270..365M}, the mass surface density dividing the Newtonian from the MONDian regime is $\Sigma_{a_0} = 275.9$~M$_\odot$~pc$^{-2}$. This is indicated in Fig.~\ref{fig:mond} by the shaded area \citep[see also fig.~7 in][]{2010A&A...523A..32K}.

Except for several very low mass GCs, dwarf galaxies with stellar masses below $\sim 10^8$~M$_\odot$ are the only objects residing deep in the MONDian regime ($a<a_0$). Apparently, the limit $\Sigma_{a_0}$ is not connected to a change in the structural properties of the galaxies, as the sequence of increasing mass surface density continues up to $M_\star \sim 10^{10}$~M$_\odot$. It is, however, interesting to note that these low mass dwarf galaxies are the only objects which exhibit very high dynamical mass-to-light ratios \citep[e.g.][]{2008MNRAS.386..864D, 2008MNRAS.389.1924F}. \citet{2010A&A...523A..32K} discuss this phenomenon in a cosmological context, and conclude that the combination of high dynamical mass-to-light ratios with $a<a_0$ is natural in a MONDian universe.

\section{Summary and conclusions}
We showed that scaling relations of dynamically hot stellar systems can be studied over a remarkable range in parameter space. Besides giant elliptical galaxies and GCs, we included -- for the first time -- large samples of cEs, UCDs, dwarf elliptical galaxies of nearby galaxy clusters, and Local Group ultra-faint dwarf spheroidals. In this way, we achieved a smooth sampling of the $R_{\mathrm{eff}}$--$M_V$ plane, the $R_{\mathrm{eff}}$--$M_\star$ plane, and the $\Sigma_{\mathrm{eff}}$--$M_\star$ plane over ten orders of magnitude in stellar mass and five orders of magnitude in effective radius.

One of the main features in the $R_{\mathrm{eff}}$--$M_V$ plane is the almost constant effective radius of galaxies in a certain magnitude range ($-20 \la M_V \la -12$~mag). Under the assumption that this is true for galaxies in all local galaxy clusters, we used this feature to determine the distances to Hydra\,I, Centaurus, Virgo, Antlia and Fornax, for which the structural parameters of a sufficiently large number of galaxies in the right magnitude range are available (Sect.~\ref{sec:distind}). It turns out that the distance estimations are in good agreement with the distances obtained with other methods, although the values derived here have rather large errors.

Star clusters and galaxies form two different families, well distinguishable in terms of their structural properties size and mass surface density. However, a closer look at the $R_{\mathrm{eff}}$--$M_\star$ and the $\Sigma_{\mathrm{eff}}$--$M_\star$ planes reveals some intriguing common features (see Fig.~\ref{fig:sizemassdiag} and Fig.~\ref{fig:surfacedensity}). For example, the similar size--mass relation of giant elliptical galaxies, cEs, UCDs, nuclei of dE,Ns and NCs, which sharply defines a maximum stellar mass for a given half-light radius (Eq.~(\ref{eq:sizemass})), translating into a maximum mass surface density, as given by Eq.~(\ref{eq:thresdensity}), creating a 'zone of avoidance' beyond these limits. Not only local early-type galaxies do not appear beyond these limits, but also most of the high-redshift galaxies ($1\la z \la 2$) presented here coincide with their local counterparts in the size/surface density--mass plane (see Fig.~\ref{fig:limit}). Given the uncertain estimates of stellar masses and effective radii of some of the high-$z$ galaxies \citep[e.g.][]{2008ApJ...677L...5V}, they might already at this time be consistent with the $z \sim 0$ objects, or they might evolve onto the relations observed for the local galaxies via subsequent merging events (Sect.~\ref{sec:zone}). However, present data on the structural parameters and stellar masses of high-$z$ ETGs are still rather limited. High quality data for a larger number of ETGs are required to be able to reach a definite conclusion on whether or not these objects are structurally different from their local counterparts.

The slope of the size--mass relation for giant elliptical galaxies is consistent with $r \propto M_\star$. This slope is predicted by the virial theorem for major galaxy mergers (see Sect.~\ref{sec:zone}). UCDs, NCs and nuclei of dE,Ns, on the other hand, can also be created by merger events \citep[e.g.][]{1998MNRAS.300..200K, 2005ApJ...618..237W, 2008ApJ...681.1136C}. Depending on the actual cluster orbital energy, the size--mass relation for merged stellar clusters is $R \propto M^{\beta}$, with $0.5 < \beta < 1$ (Merritt, unpublished\footnote{see also presentation at ESO workshop 'Central Massive Objects: The Stellar Nuclei - Black Hole Connection', \texttt{http://www.eso.org/sci/meetings/cmo2010/Presentations \linebreak /Day2/Merritt-rev.pdf}}), well in agreement with the observed slopes of $R_{\mathrm{eff}} \propto M^{3/5}$ or $R_{\mathrm{eff}} \propto M^{4/5}$. If regarding the size--mass relation as the consequence of a maximum possible stellar mass density, this might tell us something about how stars can be distributed in unrelaxed stellar systems (the median two-body relaxation time is longer than a Hubble time for these objects).

In this study we examined the mean mass surface density within the half-light radius. This radius is for giant elliptical galaxies at least 20 times larger than for star clusters. Hence, we probe regions of very different size. A better measure would be the surface density in the very central part of the particular object. \citet{2010MNRAS.401L..19H} determined this quantity for a variety of stellar systems and indeed found a maximum central stellar surface density of $\Sigma_{\mathrm{max}} \sim 10^{11}$~M$_\odot$~kpc$^{-2}$. They concluded that feedback from massive stars likely accounts for the observed $\Sigma_{\mathrm{max}}$. Unfortunately, the central density is not available with a sufficient completeness for all the stellar systems presented in this study.

The scaling relations presented here also allow to study possible formation and evolutionary scenarios, in particular those of compact stellar systems like UCDs. In contrast to usual globular clusters, these enigmatic objects exhibit a size--mass relation and enhanced mass-to-light ratios \citep[e.g.][]{2008A&A...487..921M}. Their location in the $\Sigma_{\mathrm{eff}}$--$M_\star$ plane (Fig.~\ref{fig:surfacedensity}) indicates that they are also dynamically distinct from globular clusters. With a two-body relaxation time longer than a Hubble time, they are more closely related to galaxy-like stellar systems than to regular star clusters, which have undergone considerable dynamical evolution.

In order to better understand the similarities and differences between galaxies and star clusters, and also within each family of objects, one would have to measure homogeneously the line of sight velocity dispersion along with the structural and photometric parameters (size, luminosity or surface brightness) for all of those objects. Such data would allow to explore different FP relations \citep[e.g. the Faber-Jackson relation,][]{1976ApJ...204..668F}, or the phase-space density \citep[e.g.][]{2005ApJ...618..237W, 2007ApJ...663..948G} of the entire spectrum of dynamically hot stellar systems. First studies aiming in this direction revealed interesting possible connections between galaxies and star clusters (e.g. \citealt*{2006ApJ...642L..37Z, 2006ApJ...638..725Z}; \citealt{2008MNRAS.389.1924F}, \citealt{2011arXiv1101.2460F}; \citealt*{2011ApJ...727..116Z}). However, all of these studies are lacking in large samples of low mass dwarf elliptical/spheroidal galaxies ($M_\star \la 10^9$~M$_\odot$), since it is still challenging (or even impossible) to obtain accurate velocity dispersions for such low surface brightness objects. This will be a promising science case for future ground- and space-based telescopes like the E-ELT or the JWST.

\section*{Acknowledgments}
We thank the anonymous referee for suggestions to enrich and improve the manuscript. IM acknowledges support through DFG grant BE1091/13-1. We thank Eric Emsellem, Pavel Kroupa and Steffen Mieske for helpful discussions and comments.

\bibliographystyle{mn2e}

\bibliography{hot_stellar_systems_v2}

\begin{thebibliography}{}
\bibitem[\protect\citeauthoryear{{Bekki}, {Couch}, {Drinkwater} \&
  {Gregg}}{{Bekki} et~al.}{2001}]{2001ApJ...557L..39B}
{Bekki} K.,  {Couch} W.~J.,  {Drinkwater} M.~J.,    {Gregg} M.~D.,  2001,
  \apjl, 557, L39

\bibitem[\protect\citeauthoryear{{Bekki}, {Couch}, {Drinkwater} \&
  {Shioya}}{{Bekki} et~al.}{2003}]{2003MNRAS.344..399B}
{Bekki} K.,  {Couch} W.~J.,  {Drinkwater} M.~J.,    {Shioya} Y.,  2003, \mnras,
  344, 399

\bibitem[\protect\citeauthoryear{{Belokurov}, {Zucker}, {Evans}, {Kleyna},
  {Koposov}, {Hodgkin}, {Irwin}, {Gilmore}, {Wilkinson}, {Fellhauer},
  {Bramich}, {Hewett}, {Vidrih}, {De Jong}, {Smith}, {Rix}, {Bell}, {Wyse},
  {Newberg}, {Mayeur}, {Yanny}, {Rockosi}, {Gnedin},  {Schneider},  {Beers},  {Barentine},  {Brewington},  {Brinkmann},  {Harvanek},  {Kleinman},  {Krzesinski}, {Long},  {Nitta},    {Snedden}}{{Belokurov} et~al.}{2007}]{2007ApJ...654..897B}
{Belokurov} V. et~al.,  2007, \apj, 654, 897

\bibitem[\protect\citeauthoryear{{Belokurov}, {Walker}, {Evans}, {Gilmore},
  {Irwin}, {Mateo}, {Mayer}, {Olszewski}, {Bechtold} \&
  {Pickering}}{{Belokurov} et~al.}{2009}]{2009MNRAS.397.1748B}
{Belokurov} V. et~al., 2009, \mnras, 397, 1748


\bibitem[\protect\citeauthoryear{{Belokurov}, {Walker}, {Evans}, {Gilmore},
  {Irwin}, {Just}, {Koposov}, {Mateo}, {Olszewski}, {Watkins} \&
  {Wyrzykowski}}{{Belokurov} et~al.}{2010}]{2010ApJ...712L.103B}
{Belokurov} V. et~al.,  2010, \apjl, 712, L103
  

\bibitem[\protect\citeauthoryear{{Bender}, {Burstein} \& {Faber}}{{Bender}
  et~al.}{1992}]{1992ApJ...399..462B}
{Bender} R.,  {Burstein} D.,    {Faber} S.~M.,  1992, \apj, 399, 462

\bibitem[\protect\citeauthoryear{{Bender}, {Burstein} \& {Faber}}{{Bender}
  et~al.}{1993}]{1993ApJ...411..153B}
{Bender} R.,  {Burstein} D.,    {Faber} S.~M.,  1993, \apj, 411, 153

\bibitem[\protect\citeauthoryear{{Bernardi}, {Sheth}, {Annis}, {Burles},
  {Eisenstein}, {Finkbeiner}, {Hogg}, {Lupton}, {Schlegel}, {SubbaRao},
  {Bahcall}, {Blakeslee}, {Brinkmann}, {Castander}, {Connolly}, {Csabai},
  {Doi}, {Fukugita}, {Frieman}, {Heckman},  {Hennessy},  {Ivezi{\'c}},
  {Knapp},  {Lamb},  {McKay},  {Munn},  {Nichol},
  {Okamura},  {Schneider},  {Thakar},  {York}}{{Bernardi} et~al.}{2003}]{2003AJ....125.1866B}
{Bernardi} M. et~al.,  2003, \aj, 125, 1866

\bibitem[\protect\citeauthoryear{{Binney} \& {Merrifield}}{{Binney} \&
  {Merrifield}}{1998}]{1998gaas.book.....B}
{Binney} J.,  {Merrifield} M.,  1998, {Galactic astronomy}

\bibitem[\protect\citeauthoryear{{Blakeslee}, {Jord{\'a}n}, {Mei},
  {C{\^o}t{\'e}}, {Ferrarese}, {Infante}, {Peng}, {Tonry} \&
  {West}}{{Blakeslee} et~al.}{2009}]{2009ApJ...694..556B}
{Blakeslee} J.~P. et~al.,  2009, \apj, 694, 556

\bibitem[\protect\citeauthoryear{{B{\"o}ker}, {Sarzi}, {McLaughlin}, {van der
  Marel}, {Rix}, {Ho} \& {Shields}}{{B{\"o}ker}
  et~al.}{2004}]{2004AJ....127..105B}
{B{\"o}ker} T.,  {Sarzi} M.,  {McLaughlin} D.~E.,  {van der Marel} R.~P.,
  {Rix} H.,  {Ho} L.~C.,    {Shields} J.~C.,  2004, \aj, 127, 105

\bibitem[\protect\citeauthoryear{{Brosche}}{{Brosche}}{1973}]{1973A&A....23..2%
59B}
{Brosche} P.,  1973, \aap, 23, 259

\bibitem[\protect\citeauthoryear{{Bruzual} \& {Charlot}}{{Bruzual} \&
  {Charlot}}{2003}]{2003MNRAS.344.1000B}
{Bruzual} G.,  {Charlot} S.,  2003, \mnras, 344, 1000

\bibitem[\protect\citeauthoryear{{Burstein}, {Bender}, {Faber} \&
  {Nolthenius}}{{Burstein} et~al.}{1997}]{1997AJ....114.1365B}
{Burstein} D.,  {Bender} R.,  {Faber} S.,    {Nolthenius} R.,  1997, \aj, 114,
  1365

\bibitem[\protect\citeauthoryear{{Capuzzo-Dolcetta} \&
  {Miocchi}}{{Capuzzo-Dolcetta} \& {Miocchi}}{2008}]{2008ApJ...681.1136C}
{Capuzzo-Dolcetta} R.,  {Miocchi} P.,  2008, \apj, 681, 1136

\bibitem[\protect\citeauthoryear{{Chilingarian} \& {Bergond}}{{Chilingarian} \&
  {Bergond}}{2010}]{2010MNRAS.405L..11C}
{Chilingarian} I.~V.,  {Bergond} G.,  2010, \mnras, 405, L11

\bibitem[\protect\citeauthoryear{{Chilingarian} \& {Mamon}}{{Chilingarian} \&
  {Mamon}}{2008}]{2008MNRAS.385L..83C}
{Chilingarian} I.~V.,  {Mamon} G.~A.,  2008, \mnras, 385, L83

\bibitem[\protect\citeauthoryear{{Chilingarian}, {Cayatte}, {Durret}, {Adami},
  {Balkowski}, {Chemin}, {Lagan{\'a}} \& {Prugniel}}{{Chilingarian}
  et~al.}{2008}]{2008A&A...486...85C}
{Chilingarian} I.~V.,  {Cayatte} V.,  {Durret} F.,  {Adami} C.,  {Balkowski}
  C.,  {Chemin} L.,  {Lagan{\'a}} T.~F.,    {Prugniel} P.,  2008, \aap, 486, 85

\bibitem[\protect\citeauthoryear{{Cimatti}, {Cassata}, {Pozzetti}, {Kurk},
  {Mignoli}, {Renzini}, {Daddi}, {Bolzonella}, {Brusa}, {Rodighiero},
  {Dickinson}, {Franceschini}, {Zamorani}, {Berta}, {Rosati} \&
  {Halliday}}{{Cimatti} et~al.}{2008}]{2008A&A...482...21C}
{Cimatti} A. et~al.,  2008, \aap, 482, 21

\bibitem[\protect\citeauthoryear{{Collins}, {Chapman}, {Irwin}, {Martin},
  {Ibata}, {Zucker}, {Blain}, {Ferguson}, {Lewis}, {McConnachie} \&
  {Pe{\~n}arrubia}}{{Collins} et~al.}{2010}]{2010MNRAS.407.2411C}
{Collins} M.~L.~M. et~al.,  2010, \mnras, 407, 2411

\bibitem[\protect\citeauthoryear{{C{\^o}t{\'e}}, {Piatek}, {Ferrarese},
  {Jord{\'a}n}, {Merritt}, {Peng}, {Ha{\c s}egan}, {Blakeslee}, {Mei}, {West},
  {Milosavljevi{\'c}} \& {Tonry}}{{C{\^o}t{\'e}}
  et~al.}{2006}]{2006ApJS..165...57C}
{C{\^o}t{\'e}} P. et~al.,  2006, \apjs, 165, 57

\bibitem[\protect\citeauthoryear{{C{\^o}t{\'e}}, {Ferrarese}, {Jord{\'a}n},
  {Blakeslee}, {Chen}, {Infante}, {Merritt}, {Mei}, {Peng}, {Tonry}, {West} \&
  {West}}{{C{\^o}t{\'e}} et~al.}{2007}]{2007ApJ...671.1456C}
{C{\^o}t{\'e}} P. et~al.,  2007, \apj, 671, 1456

\bibitem[\protect\citeauthoryear{{Dabringhausen}, {Hilker} \&
  {Kroupa}}{{Dabringhausen} et~al.}{2008}]{2008MNRAS.386..864D}
{Dabringhausen} J.,  {Hilker} M.,    {Kroupa} P.,  2008, \mnras, 386, 864

\bibitem[\protect\citeauthoryear{{Damjanov}, {McCarthy}, {Abraham},
  {Glazebrook}, {Yan}, {Mentuch}, {Le Borgne}, {Savaglio}, {Crampton},
  {Murowinski}, {Juneau}, {Carlberg}, {J{\o}rgensen}, {Roth}, {Chen} \&
  {Marzke}}{{Damjanov} et~al.}{2009}]{2009ApJ...695..101D}
{Damjanov} I. et~al.,  2009, \apj, 695, 101

\bibitem[\protect\citeauthoryear{{de Jong}, {Martin}, {Rix}, {Smith}, {Jin} \&
  {Macci{\`o}}}{{de Jong} et~al.}{2010}]{2010ApJ...710.1664D}
{de Jong} J.~T.~A.,  {Martin} N.~F.,  {Rix} H.,  {Smith} K.~W.,  {Jin} S.,
  {Macci{\`o}} A.~V.,  2010, \apj, 710, 1664
  
\bibitem[\protect\citeauthoryear{{Dirsch}, {Richtler} \& {Bassino}}{{Dirsch}
  et~al.}{2003}]{2003A&A...408..929D}
{Dirsch} B.,  {Richtler} T.,    {Bassino} L.~P.,  2003, \aap, 408, 929

\bibitem[\protect\citeauthoryear{{Djorgovski} \& {Davis}}{{Djorgovski} \&
  {Davis}}{1987}]{1987ApJ...313...59D}
{Djorgovski} S.,  {Davis} M.,  1987, \apj, 313, 59

\bibitem[\protect\citeauthoryear{{Drinkwater}, {Gregg}, {Hilker}, {Bekki},
  {Couch}, {Ferguson}, {Jones} \& {Phillipps}}{{Drinkwater}
  et~al.}{2003}]{2003Natur.423..519D}
{Drinkwater} M.~J.,  {Gregg} M.~D.,  {Hilker} M.,  {Bekki} K.,  {Couch} W.~J.,
  {Ferguson} H.~C.,  {Jones} J.~B.,    {Phillipps} S.,  2003, \nat, 423, 519

\bibitem[\protect\citeauthoryear{{Evstigneeva}, {Drinkwater}, {Peng}, {Hilker},
  {De Propris}, {Jones}, {Phillipps}, {Gregg} \& {Karick}}{{Evstigneeva}
  et~al.}{2008}]{2008AJ....136..461E}
{Evstigneeva} E.~A. et~al.,  2008, \aj, 136, 461

\bibitem[\protect\citeauthoryear{{Faber}}{{Faber}}{1973}]{1973ApJ...179..423F}
{Faber} S.~M.,  1973, \apj, 179, 423

\bibitem[\protect\citeauthoryear{{Faber} \& {Jackson}}{{Faber} \&
  {Jackson}}{1976}]{1976ApJ...204..668F}
{Faber} S.~M.,  {Jackson} R.~E.,  1976, \apj, 204, 668

\bibitem[\protect\citeauthoryear{{Fellhauer} \& {Kroupa}}{{Fellhauer} \&
  {Kroupa}}{2002}]{2002MNRAS.330..642F}
{Fellhauer} M.,  {Kroupa} P.,  2002, \mnras, 330, 642

\bibitem[\protect\citeauthoryear{{Ferrarese}, {C{\^o}t{\'e}}, {Jord{\'a}n},
  {Peng}, {Blakeslee}, {Piatek}, {Mei}, {Merritt}, {Milosavljevi{\'c}}, {Tonry}
  \& {West}}{{Ferrarese} et~al.}{2006}]{2006ApJS..164..334F}
{Ferrarese} L. et~al.,  2006, \apjs, 164, 334

\bibitem[\protect\citeauthoryear{{Forbes}, {Spitler}, {Graham}, {Foster}, {Hau}
  \& {Benson}}{{Forbes} et~al.}{2011}]{2011arXiv1101.2460F}
{Forbes} D.,  {Spitler} L.,  {Graham} A.,  {Foster} C.,  {Hau} G.,    {Benson}
  A.,  2011, arXiv:1101.2460
    
\bibitem[\protect\citeauthoryear{{Forbes}, {Lasky}, {Graham} \&
  {Spitler}}{{Forbes} et~al.}{2008}]{2008MNRAS.389.1924F}
{Forbes} D.~A.,  {Lasky} P.,  {Graham} A.~W.,    {Spitler} L.,  2008, \mnras,
  389, 1924

\bibitem[\protect\citeauthoryear{{Fukugita}, {Shimasaku} \&
  {Ichikawa}}{{Fukugita} et~al.}{1995}]{1995PASP..107..945F}
{Fukugita} M.,  {Shimasaku} K.,    {Ichikawa} T.,  1995, \pasp, 107, 945

\bibitem[\protect\citeauthoryear{{Gieles}, {Baumgardt}, {Heggie} \&
  {Lamers}}{{Gieles} et~al.}{2010}]{2010MNRAS.408L..16G}
{Gieles} M.,  {Baumgardt} H.,  {Heggie} D.~C.,    {Lamers} H.~J.~G.~L.~M.,
  2010, \mnras, 408, L16
    
\bibitem[\protect\citeauthoryear{{Gilmore}, {Wilkinson}, {Wyse}, {Kleyna},
  {Koch}, {Evans} \& {Grebel}}{{Gilmore} et~al.}{2007}]{2007ApJ...663..948G}
{Gilmore} G.,  {Wilkinson} M.~I.,  {Wyse} R.~F.~G.,  {Kleyna} J.~T.,  {Koch}
  A.,  {Evans} N.~W.,    {Grebel} E.~K.,  2007, \apj, 663, 948
  
\bibitem[\protect\citeauthoryear{{Graham} \& {Driver}}{{Graham} \&
  {Driver}}{2005}]{2005PASA...22..118G}
{Graham} A.~W.,  {Driver} S.~P.,  2005, Publications of the Astronomical
  Society of Australia, 22, 118
  
\bibitem[\protect\citeauthoryear{{Graham} \& {Guzm{\'a}n}}{{Graham} \&
  {Guzm{\'a}n}}{2003}]{2003AJ....125.2936G}
{Graham} A.~W.,  {Guzm{\'a}n} R.,  2003, \aj, 125, 2936

\bibitem[\protect\citeauthoryear{{Grebel}, {Gallagher} III \&
  {Harbeck}}{{Grebel} et~al.}{2003}]{2003AJ....125.1926G}
{Grebel} E.~K.,  {Gallagher} III J.~S.,    {Harbeck} D.,  2003, \aj, 125, 1926

\bibitem[\protect\citeauthoryear{{Hilker}, {Mieske} \& {Infante}}{{Hilker}
  et~al.}{2003}]{2003A&A...397L...9H}
{Hilker} M.,  {Mieske} S.,    {Infante} L.,  2003, \aap, 397, L9

\bibitem[\protect\citeauthoryear{{Hopkins}, {Bundy}, {Murray}, {Quataert},
  {Lauer} \& {Ma}}{{Hopkins} et~al.}{2009}]{2009MNRAS.398..898H}
{Hopkins} P.~F.,  {Bundy} K.,  {Murray} N.,  {Quataert} E.,  {Lauer} T.~R.,
  {Ma} C.,  2009, \mnras, 398, 898

\bibitem[\protect\citeauthoryear{{Hopkins}, {Murray}, {Quataert} \&
  {Thompson}}{{Hopkins} et~al.}{2010}]{2010MNRAS.401L..19H}
{Hopkins} P.~F.,  {Murray} N.,  {Quataert} E.,    {Thompson} T.~A.,  2010,
  \mnras, 401, L19

\bibitem[\protect\citeauthoryear{{Irwin}, {Belokurov}, {Evans}, {Ryan-Weber},
  {de Jong}, {Koposov}, {Zucker}, {Hodgkin}, {Gilmore}, {Prema}, {Hebb},
  {Begum}, {Fellhauer}, {Hewett}, {Kennicutt} Jr., {Wilkinson}, {Bramich},
  {Vidrih}, {Rix}, {Beers}, {Barentine},  {Brewington},
  {Harvanek},  {Krzesinski},  {Long},  {Nitta},  {Snedden}}{{Irwin} et~al.}{2007}]{2007ApJ...656L..13I}
{Irwin} M.~J. et~al., 2007, \apjl, 656, L13

\bibitem[\protect\citeauthoryear{{Jord{\'a}n}, {Peng}, {Blakeslee},
  {C{\^o}t{\'e}}, {Eyheramendy}, {Ferrarese}, {Mei}, {Tonry} \&
  {West}}{{Jord{\'a}n} et~al.}{2009}]{2009ApJS..180...54J}
{Jord{\'a}n} A. et~al., 2009, \apjs, 180, 54

\bibitem[\protect\citeauthoryear{{Jordi}, {Grebel} \& {Ammon}}{{Jordi}
  et~al.}{2006}]{2006A&A...460..339J}
{Jordi} K.,  {Grebel} E.~K.,    {Ammon} K.,  2006, \aap, 460, 339

\bibitem[\protect\citeauthoryear{{Kalirai}, {Beaton}, {Geha}, {Gilbert},
  {Guhathakurta}, {Kirby}, {Majewski}, {Ostheimer}, {Patterson} \&
  {Wolf}}{{Kalirai} et~al.}{2010}]{2010ApJ...711..671K}
{Kalirai} J.~S. et~al.,  2010, \apj, 711, 671

\bibitem[\protect\citeauthoryear{{Kissler-Patig}, {Jord{\'a}n} \&
  {Bastian}}{{Kissler-Patig} et~al.}{2006}]{2006A&A...448.1031K}
{Kissler-Patig} M.,  {Jord{\'a}n} A.,    {Bastian} N.,  2006, \aap, 448, 1031

\bibitem[\protect\citeauthoryear{{Koposov}, {Belokurov}, {Evans}, {Hewett},
  {Irwin}, {Gilmore}, {Zucker}, {Rix}, {Fellhauer}, {Bell} \&
  {Glushkova}}{{Koposov} et~al.}{2008}]{2008ApJ...686..279K}
{Koposov} S. et~al.,  2008, \apj, 686, 279

\bibitem[\protect\citeauthoryear{{Kormendy}}{{Kormendy}}{1977}]{1977ApJ...218.%
.333K}
{Kormendy} J.,  1977, \apj, 218, 333

\bibitem[\protect\citeauthoryear{{Kormendy}}{{Kormendy}}{1985}]{1985ApJ...295.%
..73K}
{Kormendy} J.,  1985, \apj, 295, 73

\bibitem[\protect\citeauthoryear{{Kormendy}, {Fisher}, {Cornell} \&
  {Bender}}{{Kormendy} et~al.}{2009}]{2009ApJS..182..216K}
{Kormendy} J.,  {Fisher} D.~B.,  {Cornell} M.~E.,    {Bender} R.,  2009, \apjs,
  182, 216

\bibitem[\protect\citeauthoryear{{Kroupa}}{{Kroupa}}{1998}]{1998MNRAS.300..200%
K}
{Kroupa} P.,  1998, \mnras, 300, 200

\bibitem[\protect\citeauthoryear{{Kroupa}}{{Kroupa}}{2001}]{2001MNRAS.322..231%
K}
{Kroupa} P.,  2001, \mnras, 322, 231

\bibitem[\protect\citeauthoryear{{Kroupa}, {Famaey}, {de Boer},
  {Dabringhausen}, {Pawlowski}, {Boily}, {Jerjen}, {Forbes}, {Hensler} \&
  {Metz}}{{Kroupa} et~al.}{2010}]{2010A&A...523A..32K}
{Kroupa} P. et~al.,  2010, \aap, 523, A32+

\bibitem[\protect\citeauthoryear{{{\L}okas}, {Mamon} \& {Prada}}{{{\L}okas}
  et~al.}{2005}]{2005MNRAS.363..918L}
{{\L}okas} E.~L.,  {Mamon} G.~A.,    {Prada} F.,  2005, \mnras, 363, 918

\bibitem[\protect\citeauthoryear{{Mancini}, {Daddi}, {Renzini}, {Salmi},
  {McCracken}, {Cimatti}, {Onodera}, {Salvato}, {Koekemoer}, {Aussel}, {Le
  Floc'h}, {Willott} \& {Capak}}{{Mancini} et~al.}{2010}]{2010MNRAS.401..933M}
{Mancini} C. et~al.,  2010, \mnras, 401, 933

\bibitem[\protect\citeauthoryear{{Maraston}}{{Maraston}}{2005}]{2005MNRAS.362.%
.799M}
{Maraston} C.,  2005, \mnras, 362, 799

\bibitem[\protect\citeauthoryear{{Maraston}, {Daddi}, {Renzini}, {Cimatti},
  {Dickinson}, {Papovich}, {Pasquali} \& {Pirzkal}}{{Maraston}
  et~al.}{2006}]{2006ApJ...652...85M}
{Maraston} C.,  {Daddi} E.,  {Renzini} A.,  {Cimatti} A.,  {Dickinson} M.,
  {Papovich} C.,  {Pasquali} A.,    {Pirzkal} N.,  2006, \apj, 652, 85

\bibitem[\protect\citeauthoryear{{Martin}, {de Jong} \& {Rix}}{{Martin}
  et~al.}{2008}]{2008ApJ...684.1075M}
{Martin} N.~F.,  {de Jong} J.~T.~A.,    {Rix} H.-W.,  2008, \apj, 684, 1075

\bibitem[\protect\citeauthoryear{{Martin}, {McConnachie}, {Irwin}, {Widrow},
  {Ferguson}, {Ibata}, {Dubinski}, {Babul}, {Chapman}, {Fardal}, {Lewis},
  {Navarro} \& {Rich}}{{Martin} et~al.}{2009}]{2009ApJ...705..758M}
{Martin} N.~F. et~al.,  2009, \apj, 705, 758

\bibitem[\protect\citeauthoryear{{Mateo}}{{Mateo}}{1998}]{1998ARA&A..36..435M}
{Mateo} M.~L.,  1998, \araa, 36, 435

\bibitem[\protect\citeauthoryear{{McConnachie} \& {Irwin}}{{McConnachie} \&
  {Irwin}}{2006}]{2006MNRAS.365.1263M}
{McConnachie} A.~W.,  {Irwin} M.~J.,  2006, \mnras, 365, 1263

\bibitem[\protect\citeauthoryear{{McConnachie}, {Huxor}, {Martin}, {Irwin},
  {Chapman}, {Fahlman}, {Ferguson}, {Ibata}, {Lewis}, {Richer} \&
  {Tanvir}}{{McConnachie} et~al.}{2008}]{2008ApJ...688.1009M}
{McConnachie} A.~W. et~al.,  2008, \apj, 688, 1009

\bibitem[\protect\citeauthoryear{{McLaughlin} \& {van der Marel}}{{McLaughlin}
  \& {van der Marel}}{2005}]{2005ApJS..161..304M}
{McLaughlin} D.~E.,  {van der Marel} R.~P.,  2005, \apjs, 161, 304

\bibitem[\protect\citeauthoryear{{Mei}, {Blakeslee}, {C{\^o}t{\'e}}, {Tonry},
  {West}, {Ferrarese}, {Jord{\'a}n}, {Peng}, {Anthony} \& {Merritt}}{{Mei}
  et~al.}{2007}]{2007ApJ...655..144M}
{Mei} S. et~al.,  2007, \apj, 655, 144

\bibitem[\protect\citeauthoryear{{Mieske}, {Hilker} \& {Infante}}{{Mieske}
  et~al.}{2005a}]{2005A&A...438..103M}
{Mieske} S.,  {Hilker} M.,    {Infante} L.,  2005a, \aap, 438, 103

\bibitem[\protect\citeauthoryear{{Mieske}, {Infante}, {Hilker}, {Hertling},
  {Blakeslee}, {Ben{\'{\i}}tez}, {Ford} \& {Zekser}}{{Mieske}
  et~al.}{2005b}]{2005A&A...430L..25M}
{Mieske} S.,  {Infante} L.,  {Hilker} M.,  {Hertling} G.,  {Blakeslee} J.~P.,
  {Ben{\'{\i}}tez} N.,  {Ford} H.,    {Zekser} K.,  2005b, \aap, 430, L25
  
\bibitem[\protect\citeauthoryear{{Mieske}, {Hilker}, {Infante} \& {Mendes de
  Oliveira}}{{Mieske} et~al.}{2007}]{2007A&A...463..503M}
{Mieske} S.,  {Hilker} M.,  {Infante} L.,    {Mendes de Oliveira} C.,  2007,
  \aap, 463, 503  

\bibitem[\protect\citeauthoryear{{Mieske}, {Hilker}, {Jord{\'a}n}, {Infante},
  {Kissler-Patig}, {Rejkuba}, {Richtler}, {C{\^o}t{\'e}}, {Baumgardt}, {West},
  {Ferrarese} \& {Peng}}{{Mieske} et~al.}{2008}]{2008A&A...487..921M}
{Mieske} S. et~al.,  2008, \aap, 487, 921

\bibitem[\protect\citeauthoryear{{Milgrom}}{{Milgrom}}{1983}]{1983ApJ...270..3%
65M}
{Milgrom} M.,  1983, \apj, 270, 365

\bibitem[\protect\citeauthoryear{{Misgeld}, {Mieske} \& {Hilker}}{{Misgeld}
  et~al.}{2008}]{2008A&A...486..697M}
{Misgeld} I.,  {Mieske} S.,    {Hilker} M.,  2008, \aap, 486, 697


\bibitem[\protect\citeauthoryear{{Misgeld}, {Hilker} \& {Mieske}}{{Misgeld}
  et~al.}{2009}]{2009A&A...496..683M}
{Misgeld} I.,  {Hilker} M.,    {Mieske} S.,  2009, \aap, 496, 683


\bibitem[\protect\citeauthoryear{{Murray}}{{Murray}}{2009}]{2009ApJ...691..946%
M}
{Murray} N.,  2009, \apj, 691, 946

\bibitem[\protect\citeauthoryear{{Naab}, {Johansson} \& {Ostriker}}{{Naab}
  et~al.}{2009}]{2009ApJ...699L.178N}
{Naab} T.,  {Johansson} P.~H.,    {Ostriker} J.~P.,  2009, \apjl, 699, L178

\bibitem[\protect\citeauthoryear{{Newman}, {Ellis}, {Treu} \& {Bundy}}{{Newman}
  et~al.}{2010}]{2010ApJ...717L.103N}
{Newman} A.~B.,  {Ellis} R.~S.,  {Treu} T.,    {Bundy} K.,  2010, \apjl, 717,
  L103


\bibitem[\protect\citeauthoryear{{Niederste-Ostholt}, {Belokurov}, {Evans},
  {Gilmore}, {Wyse} \& {Norris}}{{Niederste-Ostholt}
  et~al.}{2009}]{2009MNRAS.398.1771N}
{Niederste-Ostholt} M.,  {Belokurov} V.,  {Evans} N.~W.,  {Gilmore} G.,  {Wyse}
  R.~F.~G.,    {Norris} J.~E.,  2009, \mnras, 398, 1771

\bibitem[\protect\citeauthoryear{{Paturel}, {Petit}, {Prugniel}, {Theureau},
  {Rousseau}, {Brouty}, {Dubois} \& {Cambr{\'e}sy}}{{Paturel}
  et~al.}{2003}]{2003A&A...412...45P}
{Paturel} G.,  {Petit} C.,  {Prugniel} P.,  {Theureau} G.,  {Rousseau} J.,
  {Brouty} M.,  {Dubois} P.,    {Cambr{\'e}sy} L.,  2003, \aap, 412, 45

\bibitem[\protect\citeauthoryear{{Paudel} \& {Lisker}}{{Paudel} \&
  {Lisker}}{2009}]{2009AN....330..969P}
{Paudel} S.,  {Lisker} T.,  2009, Astronomische Nachrichten, 330, 969

\bibitem[\protect\citeauthoryear{{Paudel}, {Lisker}, {Kuntschner}, {Grebel} \&
  {Glatt}}{{Paudel} et~al.}{2010}]{2010MNRAS.405..800P}
{Paudel} S.,  {Lisker} T.,  {Kuntschner} H.,  {Grebel} E.~K.,    {Glatt} K.,
  2010, \mnras, 405, 800

\bibitem[\protect\citeauthoryear{{Peng}, {C{\^o}t{\'e}}, {Jord{\'a}n},
  {Blakeslee}, {Ferrarese}, {Mei}, {West}, {Merritt}, {Milosavljevi{\'c}} \&
  {Tonry}}{{Peng} et~al.}{2006}]{2006ApJ...639..838P}
{Peng} E.~W. et~al.,  2006, \apj, 639, 838

\bibitem[\protect\citeauthoryear{{Pflamm-Altenburg} \&
  {Kroupa}}{{Pflamm-Altenburg} \& {Kroupa}}{2009}]{2009MNRAS.397..488P}
{Pflamm-Altenburg} J.,  {Kroupa} P.,  2009, \mnras, 397, 488

\bibitem[\protect\citeauthoryear{{Price}, {Phillipps}, {Huxor}, {Trentham},
  {Ferguson}, {Marzke}, {Hornschemeier}, {Goudfrooij}, {Hammer}, {Tully},
  {Chiboucas}, {Smith}, {Carter}, {Merritt}, {Balcells}, {Erwin} \&
  {Puzia}}{{Price} et~al.}{2009}]{2009MNRAS.397.1816P}
{Price} J. et~al.,  2009, \mnras, 397, 1816

\bibitem[\protect\citeauthoryear{{Rossa}, {van der Marel}, {B{\"o}ker},
  {Gerssen}, {Ho}, {Rix}, {Shields} \& {Walcher}}{{Rossa}
  et~al.}{2006}]{2006AJ....132.1074R}
{Rossa} J.,  {van der Marel} R.~P.,  {B{\"o}ker} T.,  {Gerssen} J.,  {Ho}
  L.~C.,  {Rix} H.,  {Shields} J.~C.,    {Walcher} C.,  2006, \aj, 132, 1074

\bibitem[\protect\citeauthoryear{{Smith Castelli}, {Bassino}, {Richtler},
  {Cellone}, {Aruta} \& {Infante}}{{Smith Castelli}
  et~al.}{2008a}]{2008MNRAS.386.2311S}
{Smith Castelli} A.~V.,  {Bassino} L.~P.,  {Richtler} T.,  {Cellone} S.~A.,
  {Aruta} C.,    {Infante} L.,  2008a, \mnras, 386, 2311

\bibitem[\protect\citeauthoryear{{Smith Castelli}, {Faifer}, {Richtler} \&
  {Bassino}}{{Smith Castelli} et~al.}{2008b}]{2008MNRAS.391..685S}
{Smith Castelli} A.~V.,  {Faifer} F.~R.,  {Richtler} T.,    {Bassino} L.~P.,
  2008b, \mnras, 391, 685

\bibitem[\protect\citeauthoryear{{Tiret}, {Salucci}, {Bernardi}, {Maraston} \&
  {Pforr}}{{Tiret} et~al.}{2010}]{2010MNRAS.tmp.1737T}
{Tiret} O.,  {Salucci} P.,  {Bernardi} M.,  {Maraston} C.,    {Pforr} J.,
  2010, \mnras, pp 1737--+
  
\bibitem[\protect\citeauthoryear{{Tollerud}, {Bullock}, {Graves} \&
  {Wolf}}{{Tollerud} et~al.}{2011}]{2011ApJ...726..108T}
{Tollerud} E.~J.,  {Bullock} J.~S.,  {Graves} G.~J.,    {Wolf} J.,  2011, \apj,
  726, 108
  
\bibitem[\protect\citeauthoryear{{van Dokkum}, {Franx}, {Kriek}, {Holden},
  {Illingworth}, {Magee}, {Bouwens}, {Marchesini}, {Quadri}, {Rudnick},
  {Taylor} \& {Toft}}{{van Dokkum} et~al.}{2008}]{2008ApJ...677L...5V}
{van Dokkum} P.~G. et~al.,  2008, \apjl, 677, L5

\bibitem[\protect\citeauthoryear{{Walcher}, {van der Marel}, {McLaughlin},
  {Rix}, {B{\"o}ker}, {H{\"a}ring}, {Ho}, {Sarzi} \& {Shields}}{{Walcher}
  et~al.}{2005}]{2005ApJ...618..237W}
{Walcher} C.~J. et~al.,  2005, \apj, 618, 237

\bibitem[\protect\citeauthoryear{{Zaritsky}, {Zabludoff} \&
  {Gonzalez}}{{Zaritsky} et~al.}{2011}]{2011ApJ...727..116Z}
{Zaritsky} D.,  {Zabludoff} A.~I.,    {Gonzalez} A.~H.,  2011, \apj, 727, 116

\bibitem[\protect\citeauthoryear{{Zaritsky}, {Gonzalez} \&
  {Zabludoff}}{{Zaritsky} et~al.}{2006a}]{2006ApJ...642L..37Z}
{Zaritsky} D.,  {Gonzalez} A.~H.,    {Zabludoff} A.~I.,  2006a, \apjl, 642, L37

\bibitem[\protect\citeauthoryear{{Zaritsky}, {Gonzalez} \&
  {Zabludoff}}{{Zaritsky} et~al.}{2006b}]{2006ApJ...638..725Z}
{Zaritsky} D.,  {Gonzalez} A.~H.,    {Zabludoff} A.~I.,  2006b, \apj, 638, 725


\end{thebibliography}

\label{lastpage}

\end{document}